\documentclass{aa} 
\usepackage{graphicx}
\usepackage{txfonts}
\usepackage{xcolor}
\usepackage{amsmath} 
\usepackage{float}          
\usepackage{subcaption}    
\usepackage{multirow}      
\usepackage{array}         
\usepackage{longtable}
\usepackage{pdflscape}    

\usepackage{placeins}

\usepackage{hyperref} 
\hypersetup{
    colorlinks,
    linkcolor={red!50!black},
    citecolor={blue!50!black},
    urlcolor={green!80!black}
} 

\begin{document} 

   \title{CORALIE radial-velocity search for companions around evolved stars (CASCADES)}
   
   \subtitle{V. Three planetary companions and achievable precision}

   \author{P. Figueira\inst{1,2}\corrauth{pedro.figueira@iaa.es}, 
            E. Fontanet\inst{2}\email{emile.fontanet@unige.ch},
            J. P. Faria\inst{2}\email{joao.faria@unige.ch},
            M. Esseldeurs\inst{3}\email{mats.esseldeurs@kuleuven.be},
            E. Friden\inst{2}\email{erik.friden@unige.ch},
            A. Leleu\inst{2}\email{adrien.leleu@unige.ch},
            R. Luque\inst{1}\email{rluque@iaa.es}, \\
            G. Ottoni\inst{2}\email{gael.ottoni@unige.ch},
            D. S\'{e}gransan\inst{2}\email{damien.segransan@unige.ch},
            M. Stalport\inst{4,5}\email{manu.stalport@uliege.be},
            S. Tavella\inst{2,6}\email{sara.tavella@unige.ch},
            S. Udry\inst{2}\email{stephane.udry@unige.ch}
          }

   \institute{Instituto de Astrof\'{i}sica de Andaluc\'{i}a-CSIC, Glorieta de la Astronom\'{i}a s/n, 18008 Granada, Spain
        \and
        Observatoire Astronomique de l’Universit\'{e} de Gen\`{e}ve, Chemin Pegasi 51b, 1290 Versoix, Switzerland
        \and 
         Instituut voor Sterrenkunde, KU Leuven, Celestijnenlaan 200D, 3001 Leuven, Belgium
         \and
        Space sciences, Technologies and Astrophysics Research (STAR) Institute, Universit\'{e} de Li\`{e}ge, All\'{e}e du 6 Ao\^{u}t 19C, 4000 Li\`{e}ge, Belgium 
        \and
        Astrobiology Research Unit, Universit\'{e} de Li\`{e}ge, All\'{e}e du 6 Ao\^{u}t 19C, 4000 Li\`{e}ge, Belgium 
        \and 
        European Southern Observatory, Alonso de C\'{o}rdova 3107, Vitacura, Regi\'{o}n Metropolitana, Chile
             }

   \date{Received 20 March 2026; Accepted 08 June 2026}

  \abstract
   {}
   {We extend the planetary census around massive stellar hosts through a long-term campaign of high-precision radial velocity (RV) measurements on evolved stars.}
   {We analysed data acquired with the CORALIE spectrograph covering 15-18 years on HD\,125136, HD\,127195, and HD\,220218. The stellar parameters were derived through different methods for a comprehensive characterisation of each star. We then evaluated the presence of planetary signals in the RV time series using the Bayesian inference tool \texttt{kima}. Finally, we designed an observing strategy aimed at mitigating the impact of pulsations on evolved stars and tested its effectiveness on the low-luminosity red giant HD\,127195.}
   {We detected signals that have been accurately modelled by Keplerian curves in the RV data of the three stars: one on HD\,125136, two on HD\,127195, and one on HD\,220218. While the signals on the first two stars appear to be planetary in origin, the signal on the third one shows several signs of stellar activity. Therefore, we identified a planetary companion around HD\,125136 with a minimum mass of 2.26\,M$_{\mathrm{Jup}}$ on an 850\,d orbit. On HD\,127195, we identified a system composed of planets with 0.66\,M$_{\mathrm{Jup}}$ and 0.78\,M$_{\mathrm{Jup}}$ with orbital periods of 535\,d and 834\,d, respectively. 
   }
   {We detected three massive planets around two low-luminosity red giant stars in a region of the parameter space that is poorly populated in both stellar mass and planetary orbital periods. The dedicated observing campaign on HD\,127195 showcases how stellar pulsations can be efficiently averaged out to below 5\,m\,s$^{-1}$ in low-luminosity giant stars.}


   \titlerunning{Exoplanets around evolved stars}
   \authorrunning{Figueira et al. (2025)}
   \maketitle
   \nolinenumbers

\section{Introduction}

With approximately 6200 planets known\footnote{Check \href{http://exoplanet.eu/}{exoplanet.eu} for an updated value.}, the search for exoplanets is now mainstream science. While the search for planets around low-mass stars has received considerable attention, the study of planets around stars more massive than our Sun has not. Currently, there are only $\sim$1300 planets detected around stars with a mass greater than 1.1\,M$_\odot$, which corresponds to about 20\% of the planets known.

The study of planets around stars more massive than our Sun is nonetheless of undisputed scientific relevance. Because stellar mass is a proxy for protoplanetary disk mass and lifetime, planets around massive stars provide key constraints on planet formation efficiency, migration, and atmospheric evolution. Planetary parameters are a function of stellar mass \citep[e.g.][]{2005A&A...434..343A, 2011A&A...526A..63A, 2008ApJ...673..502K} and planetary demographics features such as the radius-gap \citep{2017AJ....154..109F, 2018MNRAS.479.4786V} are also strongly dependent on the host mass, as demonstrated by \cite{2024A&A...686L...9V}. 

Unfortunately, the higher effective temperature of massive stars when compared with solar-mass stars reduces the number of photospheric lines that can be used for precise radial velocity (RV) determination. Moreover, the median rotational velocity increases with mass \citep[e.g.][]{1965Obs....85..166M}. A-type stars typically exhibit projected rotational velocities $v \sin i \gtrsim$100\,km\,s$^{-1}$, while F-type stars show a wide distribution, often reaching several tens of km\,s$^{-1}$. The survey of planets around AF main sequence stars with the HARPS spectrograph confirmed this dependence of achievable precision on stellar mass \citep{2019A&A...621A..87B}. The authors reported a median RV dispersion of 60\,m\,s$^{-1}$ for a sample with a median photon-noise uncertainty of 14\,m\,s$^{-1}$, with both quantities strongly dependent on stellar mass, spectral type, and rotational velocity.

When targeting stars in the 1.1--5\,M$_\odot$ mass range, most campaigns focus on old stars that moved out of the main sequence \citep[e.g.][]{2007A&A...472..657L}. These evolved stars are more luminous, have a larger radius, and a lower effective temperature than their main sequence counterparts; they have historically been labelled as giants in opposition to the main sequence dwarfs. The difference in observables comes from the star burning hydrogen in a shell around an inert helium core (and later, for sufficiently massive stars, igniting helium in the core) that pushes the stellar layers outwards and modifies the properties of the photosphere.

We have been pursuing a long-term RV follow-up of evolved stars, named CASCADES. The programme and first results were presented by \cite{2022A&A...657A..87O}, with three follow-up works on stellar and planetary characterisation by \cite{2022A&A...657A..88B}, \cite{2022A&A...657A..89P} and the latest by \cite{2025A&A...699A..38F}.

In this work, we focus on targets on the early phase of the red giant evolutionary stage. These stars are often called low-luminosity red giants and exhibit a comparatively lower RV jitter than their higher luminosity counterparts. We used this characteristic to our advantage to explore the planets population detectable in RV. These systems provide a crucial bridge between planet populations around solar-type stars and those forming in the higher mass, shorter lived disk environments of intermediate-mass stars.

We start with the stellar characterisation and the properties of CORALIE data in Section\,\ref{Sec:SteDat}. In Section\,\ref{Sec:RVsignals}, we study the signals present in our RVs and their nature. In Section\,\ref{Sec:PulsPrec}, we present a dedicated campaign on one of our subgiants aimed at reducing the impact of pulsations on the error budget and evaluate its efficiency. We discuss and contextualise our results in Section\,\ref{Sec:Disc} and wrap up our conclusions in Section\,\ref{Sec:Conc}.

\section{Target properties and data}\label{Sec:SteDat}

\begin{figure*}
\center
\includegraphics[width=0.8\textwidth]{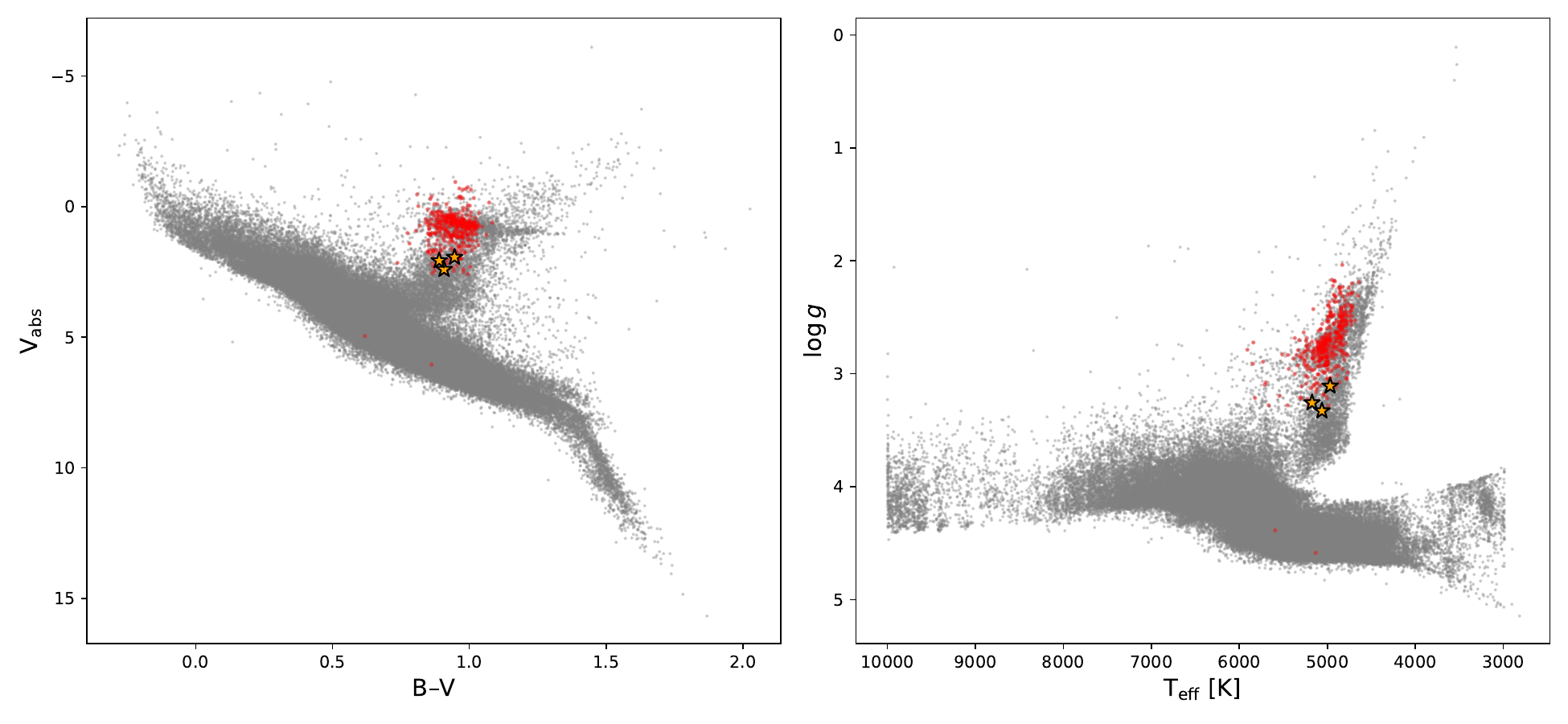}
\caption{HR (\textit{left}) and Kiel (\textit{right}) diagrams for the CASCADES stars (red dots) and the three stars analysed in this work (orange stars), against the \textit{Gaia} DR3 sample (\textit{gray dots}).}\label{Fig:CASCADES_vs_Gaia}
\centering
\end{figure*}

In Fig.\,\ref{Fig:CASCADES_vs_Gaia}, we represent the CASCADES survey stars against the sample of bright ($G$\,<\,12\,mag) nearby ($d$\,<\,200\,pc) \textit{Gaia} Data Release 3 \citep[][henceforth DR3]{2016A&A...595A...1G, 2023A&A...674A...1G} stars. We used an observational Hertzsprung-Russell (HR) diagram and a Kiel diagram to represent the evolutionary stages graphically To transform \textit{Gaia}'s $G$ into $V$, we used Table 5.9 of \textit{Gaia} DR3 documentation\footnote{\url{https://gea.esac.esa.int/archive/documentation/GDR3/Data_processing/chap_cu5pho/cu5pho_sec_photSystem/cu5pho_ssec_photRelations.html}}. For display purposes, we selected only objects with 2000\,<\,$T_{\mathrm{eff}}$\,<\,10\,000\, K, and --0.5\,mag $< B-V <$ 2.5\,mag. While in the HR diagram, the evolved stars depart from the main sequence becoming brighter and redder, in the Kiel diagram the giant branch is clearly separated toward lower $\log g$ and lower $T_{\mathrm{eff}}$ \citep[e.g.][]{1998A&A...338..161F, 2005PASJ...57...27T}.

\subsection{Target properties}

In the past, several exoplanet programmes have used absolute magnitude and colour cuts to select for subgiant stars \citep[e.g.][]{2007ApJ...665..785J}. These stars have left the main sequence, exhibiting a lower temperature and higher luminosity, but have not yet ascended into the red giant branch, showing a particularly low RV jitter for an evolved star. We selected stars from the CASCADES catalogue by applying the constraints 1.8\,$<$\,$M_V$\,$<$\,2.5\,mag and 0.78\,$<$\,$B-V$\,$<$\,1.06\,mag, following the prescription of \cite{2011ApJ...743..184W}. From the 26 stars satisfying these criteria, we identified three stars worth studying in detail: HD\,125136, HD\,127195, and HD\,220218. These targets stood out for having a sufficient number of RV measurements for detailed analysis ($\approx$\,40 epochs) and an RV scatter of several tens of m\,s$^{-1}$, compatible with the signature induced by a Jupiter-mass companion. 

For each star, we compiled the parameters published in \cite{2022A&A...657A..87O} and complemented these with both the \textit{Gaia} DR3 parameters and parameters derived with \texttt{ARIADNE} \citep{2022MNRAS.513.2719V}. \texttt{ARIADNE} is a Python package for fitting spectral energy distributions using Bayesian inference to derive fundamental stellar parameters from multi-band photometry. The data collection, parameter derivation, and resulting stellar parameters for each case are presented in Appendix\,\ref{App:SteChar}. A comparison of the results, presented in Table\,\ref{Tab:SteParams} and illustrated by Fig.\,\ref{Fig:SteParams} reveals both similarities and differences. Except for a few cases, the stellar parameters derived for a given star agree across methods within the quoted uncertainties. In the remainder of the paper, we adopted the CASCADES values, which are available for all stars and parameters.

Subgiant stars typically have R $\sim$ 1.2--3.5\,$R_\odot$ and $\log{g}\,\sim$3.7--4.1 \citep[e.g.][]{2025MNRAS.544.1186B}. The radii that we measured for our targets is larger and the surface gravity lower than those of subgiant stars. Moreover, our targets are not located on the horizontal subgiant branch in the HR diagram, but above it. We then classified these stars as low-luminosity giant stars. To the best of our knowledge, there is no prior literature providing a detailed characterisation of these stars, nor dedicated studies of planet occurrence around them.

\subsection{CORALIE spectrograph and data properties}

The CORALIE spectrograph is a twin of the ELODIE spectrograph \citep{1996A&AS..119..373B}, used to detect 51\,Peg\,b \citep{1995Natur.378..355M}. Mounted on the 1.2\,m Swiss Euler telescope at La Silla, it has been in continuous operation since 1998 \citep{2000A&A...354...99Q} and contributed to more than 200 refereed works\footnote{Data recovered with the \href{https://ui.adsabs.harvard.edu/}{The SAO Astrophysics Data System} in June 2026.}. CORALIE is an optical (381 to 681\,nm) fibre-fed high-resolution (R $\sim$ 60\,000) echelle spectrograph. In parallel with the science fibre (A), a second calibration one (B) is used to collect either skylight for background measurement, or light from a RV reference: either a Th--Ar lamp of a back-illuminated Fabry--Pérot (FP) etalon. This reference is used to measure the instrumental drift, a first-order estimation of the wavelength solution shift due to environmental changes. Since the light collected by the two fibres follows parallel and very similar optical paths, we can assume that the shift measured in B is likewise present in the stellar spectra collected through A. The instrument's drift is then corrected by subtracting from the science target RV the contemporaneous drift.

The data acquired with CORALIE are reduced with an automatic data reduction software (\texttt{DRS}), which ensures a homogeneous RV derivation. The RVs are calculated via the cross-correlation function method, as described in \cite{1996A&AS..119..373B} and expanded upon to include weighted masks \citep{2002A&A...388..632P}. Since our three stars are characterised by $T_{\mathrm{eff}} \approx 5000$ K, corresponding to early-K spectral types, we used the K5 template mask as the closest available option. Upon reduction, the data are injected into the \texttt{DACE} database and available through the dedicated interface\footnote{\url{https://dace.unige.ch/radialVelocities/?}}.

The cross-correlation function is provided with the quantities full width at half maximum (FWHM), line contrast, and bisector inverse slope \citep[BIS, see][]{2001A&A...379..279Q}, which tracks shape variations of the photospheric lines. In addition, \texttt{DACE} performs a post-processing calculation of H$\alpha$ activity index. The index is calculated over a narrow wavelength region, following the prescription and wavelength limits defined in \cite{2007A&A...469..309C} and \cite{2009A&A...495..959B}. The customary Na~{\sc i} and Ca~{\sc ii} activity indices were not used; the low chromospheric flux when compared to the strong contamination from the FP spectra make the measurements useless. 

CORALIE underwent several interventions that changed its optical properties or wavelength dispersion characteristics. As such, the time series are divided into independent datasets, treated as different instruments:

\begin{itemize}
    \item \texttt{COR98, DRS-3.3}: the original CORALIE setup \citep[see][]{1996A&AS..119..373B, 2000A&A...354...99Q};
    \item \texttt{COR07, DRS-3.4}: in April--May 2007 the fibre link and cross-disperser optics were replaced, leading to an increase in resolution and throughput. Since the double-scrambler was removed, the illumination stability properties were reduced and its contribution to the RV budget increased, leading to a slightly lower RV precision than for \texttt{COR98}. For further information see \cite{2010A&A...511A..45S};
    \item \texttt{COR14, DRS-3.8}: in November 2014 octagonal fibers were installed for improved light scrambling (without reducing throughput) and an FP was installed for simultaneous drift measurement in fibre B, replacing the Th--Ar lamp that had been used for this role, (along with being used for wavelength solution, as is customary);
    \item \texttt{COR24, DRS-3.8}: On 15 October 2024 an RV offset was introduced due to a change of the Th--Ar lamp.
\end{itemize}

The instrument has demonstrated a long-term RV precision of around 5\,m\,s$^{-1}$ \citep{2010A&A...511A..45S}.


\section{Planetary signals in RVs}\label{Sec:RVsignals}

The CORALIE RVs and associated quantities were retrieved from the \texttt{DACE} database using the \texttt{arvi} package\footnote{\url{https://github.com/j-faria/arvi}}. 
In our analysis, we discarded observations with photon noise larger than 30\,m\,s$^{-1}$ (associated with very poor weather conditions) or for which the instrument drift was not properly corrected. The latter is identified using the drift quality control (QC) flag, and a complete explanation is presented in Appendix\,\ref{App:CORALIEkws}.

\subsection{Simple statistics and general properties}\label{Sec:RVsimple}

\begin{figure*}[t]
\center
\includegraphics[width=0.95\textwidth]{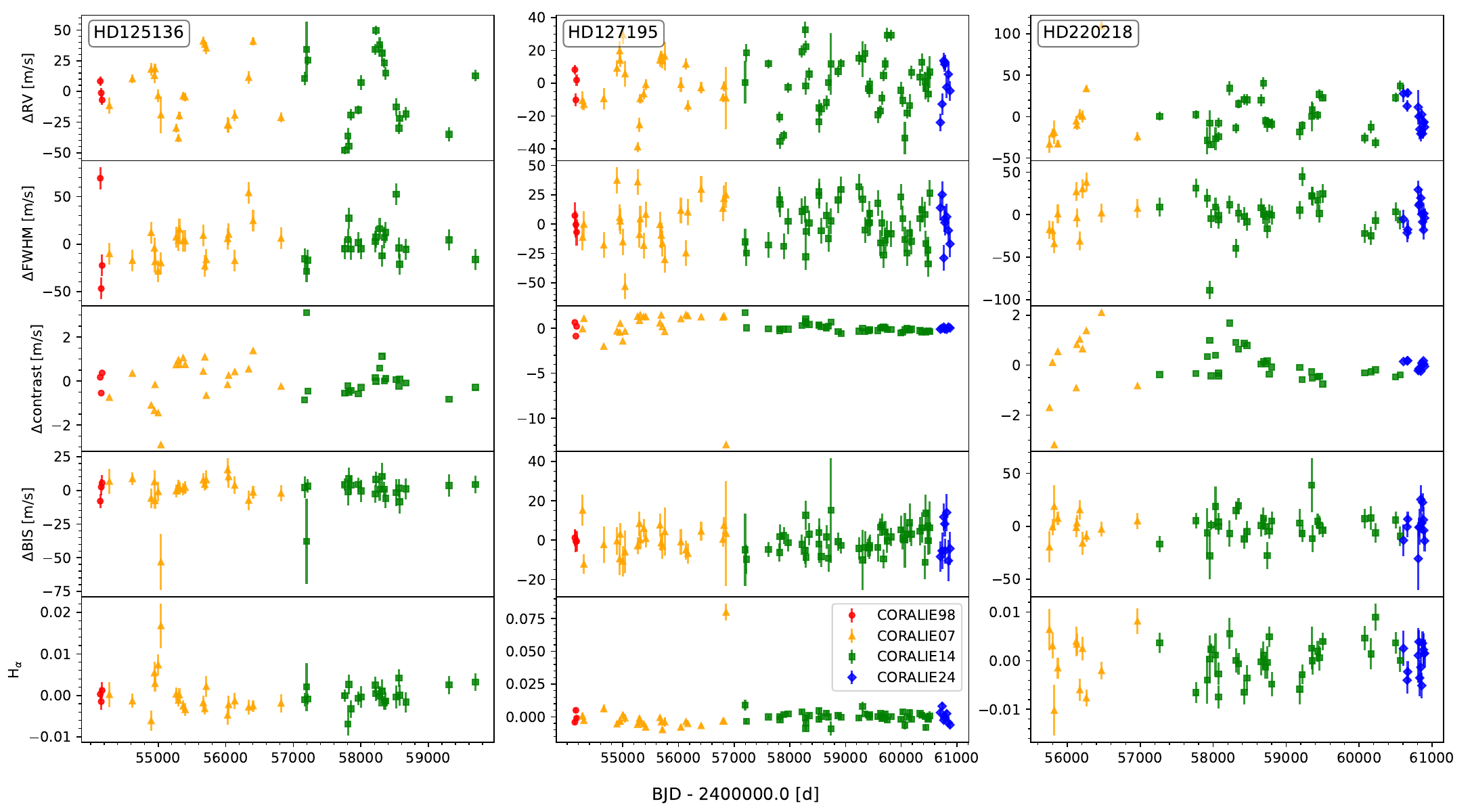}
\caption{Time series of RV and activity indicators: FWHM, contrast, BIS, and H$\alpha$ (\textit{top to bottom}) for the stars HD\,125136 (\textit{left}), HD\,127195 (\textit{center}), and HD\,220218 (\textit{right}). The different CORALIE datasets are represented as different colours and symbols.}\label{Fig:timeseries}
\centering
\end{figure*}

The stars were observed for a total time span of 15--18 years, and each star's time series, shown in Fig.\,\ref{Fig:timeseries}, contains between 45 and 81 points. For each star and dataset we provide simple summary statistics in Table\,\ref{Tab:RVstimeseries}. The average (photon-noise) RV uncertainty of each star's full dataset is 5--8\,m\,s$^{-1}$, comparable to or larger than the expected instrumental stability. However, the scatter, measured through the weighted root mean square (rms) is 16--28 m\,s$^{-1}$, in excess of both formal uncertainties and the instrument stability level. The difference between these quantities points towards an additional signal or unaccounted source of noise.

\setlength{\extrarowheight}{3pt}

\begin{table}
\centering
\caption{Simple statistics for the different RV datasets acquired.}
\label{Tab:RVstimeseries}
\resizebox{\columnwidth}{!}{%

\begin{tabular}{ll c c c  c c c }
\hline  \hline
 \multirow{2}{*}{Star} & \multirow{2}{*}{Dataset} & $\Delta t$ & \multicolumn{2}{c}{$N_{\mathrm{points}}$} & avg. err & w. rms  \\
    &   &  \tiny{[d]} &  \tiny{QC$_{\mathrm{pass}}$} & \tiny{QC$_{\mathrm{fail}}$} & \multicolumn{2}{c}{\tiny{[m\,s$^{-1}$}]}  \\[2pt] 
\hline 

 \multirow{4}{*}{HD\,125136} & full set & 5552.9 & 45 & 4 & 5.4 & 27.4 \\ 
& \tiny{\texttt{COR98}} & 27.0 & 3 & 0 & 4.0 & 6.5 \\ 
& \tiny{\texttt{COR07}} & 2541.0 & 21 & 1 & 4.9 & 26.6 \\ 
& \tiny{\texttt{COR14}} & 2527.0 & 21 & 3 & 6.0 & 31.1 \\[4pt] 

 \multirow{5}{*}{HD\,127195} & full set & 6737.6 & 81 & 5 & 5.2 & 15.6 \\ 
& \tiny{\texttt{COR98}} & 31.0 & 3 & 0 & 3.5 & 7.4 \\ 
& \tiny{\texttt{COR07}} & 2576.9 & 24 & 1 & 5.0 & 15.5 \\ 
& \tiny{\texttt{COR14}} & 3316.9 & 46 & 3 & 5.3 & 16.5 \\ 
& \tiny{\texttt{COR24}} & 172.6 & 8 & 1 & 5.8 & 13.1 \\[4pt]

\multirow{4}{*}{HD\,220218} & full set & 5141.0 & 54 & 4 & 7.7 & 28.2 \\ 
& \tiny{\texttt{COR07}} & 1206.7 & 11 & 0 & 6.9 & 43.9 \\ 
& \tiny{\texttt{COR14}} & 3295.0 & 31 & 4 & 7.7 & 21.8 \\ 
& \tiny{\texttt{COR24}} & 290.2 & 12 & 0 & 8.7 & 16.7 \\[4pt] 

 \hline

\end{tabular}
}
\end{table}

\setlength{\extrarowheight}{2pt}

\subsection{Periodic signals and correlation with activity indicators}

\begin{figure*}[t]
\center
\includegraphics[width=0.95\textwidth]{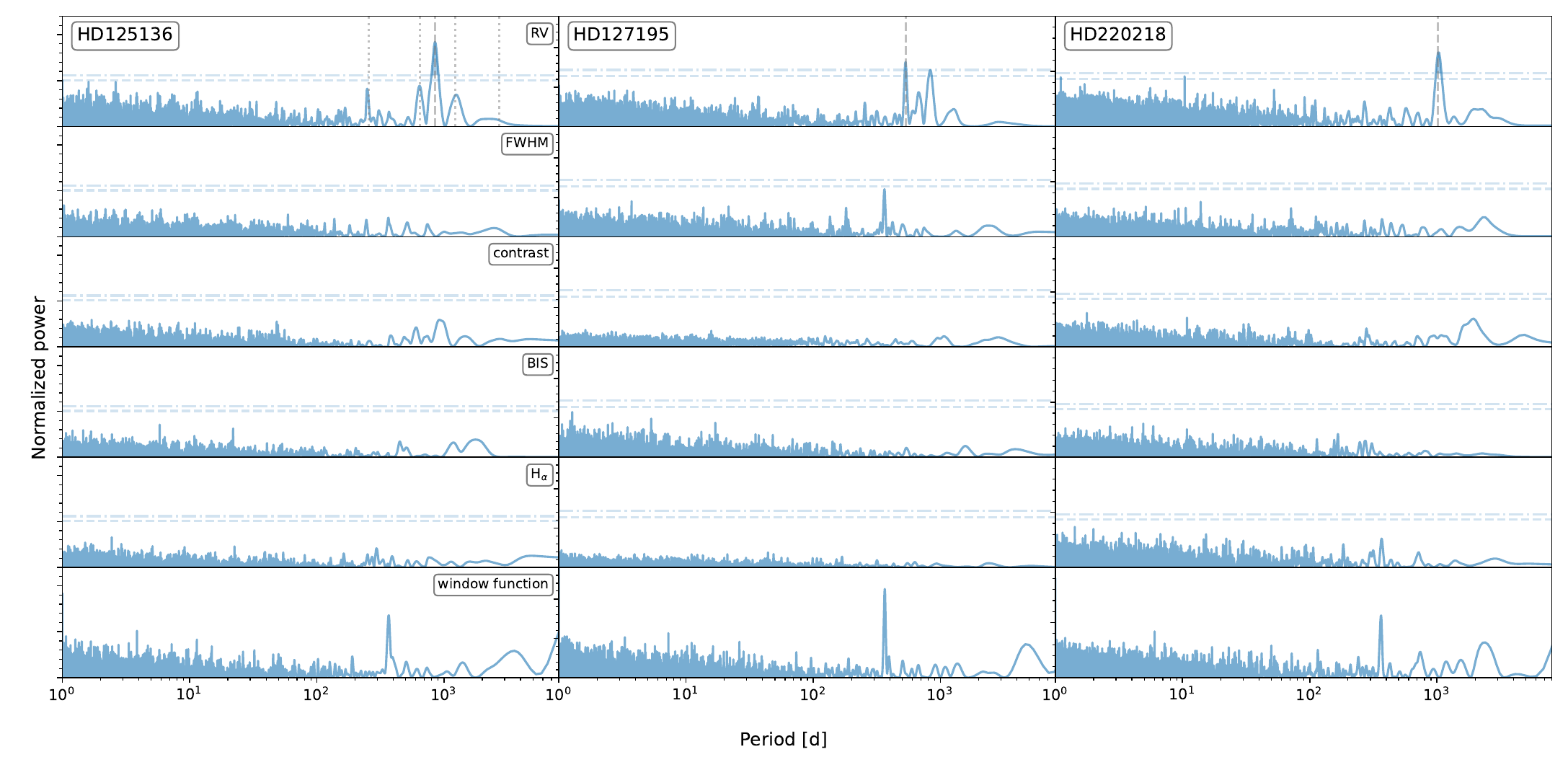}
\caption{GLS periodogram of RV, FWHM, contrast, BIS, H$\alpha$, and of the window function (\textit{top to bottom}) for the stars HD\,125136 (\textit{left}), HD\,127195 (\textit{center}), and HD\,220218 (\textit{right}). The FAP levels of 1 and 0.1\% are shown as horizontal lines. The highest significance peaks are show as dashed vertical lines and the aliases for HD\,125136 are shown as dotted vertical lines.}\label{Fig:gls}
\centering
\end{figure*}

We computed the generalised Lomb-Scargle periodogram (GLS) as introduced by \cite{2009A&A...496..577Z} and implemented in \texttt{astropy}. To evaluate the significance of each signal we used the false-alarm probability (FAP) as defined by \cite{2008MNRAS.385.1279B}. We chose two FAP threshold probabilities of 1\% and 0.1\% and a maximum frequency of 1\,d$^{-1}$ for the GLS and FAP calculations. The GLS periodogram for the different RV sets is shown in Fig.\,\ref{Fig:gls}, along with the FAP thresholds. The window function, estimated by setting the measured signal at 1.0 \citep[see][]{2018ApJS..236...16V}, is shown in the bottom row.

Each star shows a significant periodicity at the 0.1\% FAP level: 848\,d, 533\,d, and 1039\,d. In parallel, the window function shows a clear periodicity at around one year, caused by the annual visibility window of the targets from La Silla.

In addition to the most significant peak, the periodogram of HD\,125136 shows several alias periods caused by the annual sampling window. Using the formalism described by \cite{2018ApJS..236...16V}, 256\,d and 647\,d are the typical aliases, identified with ($n$\,=\,1, $m$\,=\,1). The peak and bump at 1228\,d and 2735\,d correspond to ($n$\,=\,1, $m$\,=\,3) and ($m$\,=\,2, $n$\,=\,--1), respectively. The integer $m$ being larger than one comes from a high-order harmonic of the true frequency, and the negative $n$ comes from the peak alias crossing into the negative frequency regime and being reflected into the positive frequency range, as discussed by \cite{2018ApJS..236...16V}. This representation of the peak series as several aliases only demonstrates the presence of an underlying signal; from the frequency analysis, it is not possible to separate original and aliased periods.

HD\,127195 shows a second peak at 832\,d, reaching the 0.1\% significance threshold. We could not identify any plausible alias near to this period and, thus, we interpret this signal as real. The activity indicators FWHM, contrast, BIS, and H${\alpha}$ do not show significant periodic signals. However, for HD\,127195 the FWHM has a marginal peak near the 1\% FAP level, for a period of approximately one year. 

\begin{figure*}
\center
\includegraphics[width=\textwidth]{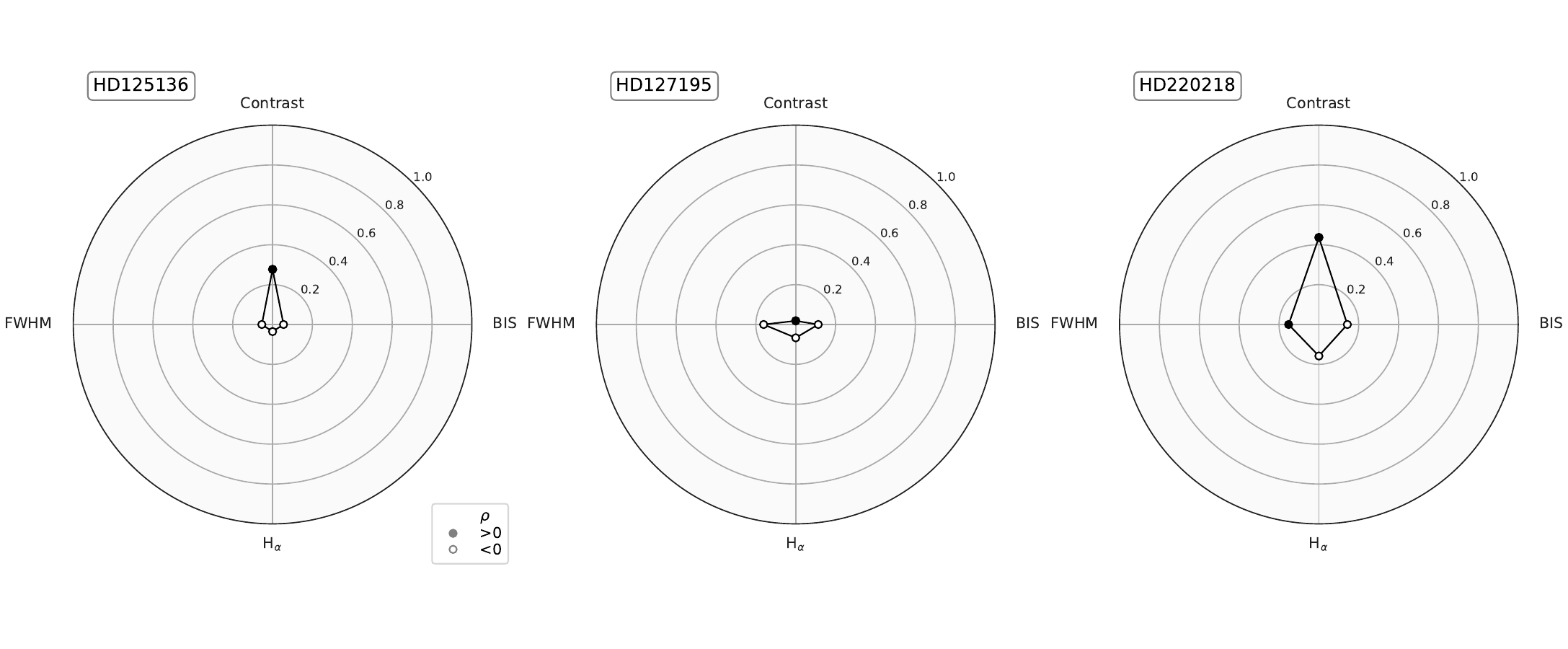}
\caption{Radar chart of the correlation between RV and each of the indicators, for the three stars studied in the paper. The filled and open dots representing positive or negative signed Pearson correlation coefficients, respectively.}\label{Fig:radarcorr}
\centering
\end{figure*}

A correlation between RVs and an activity indicator usually points towards an activity-induced RV signal \citep[e.g.][]{2013A&A...557A..93F}. We check for correlations between the RVs and the FWHM, bisector, contrast and H${\alpha}$ indicators using the Pearson Correlation coefficient. The coefficient values are shown in the radar chart in Fig\,\ref{Fig:radarcorr}, for a one-shot representation of correlations for a given star. The different indicators exhibit weak to no correlations with RV for the three stars.

The correlation coefficient for FWHM, BIS, and $H{\alpha}$ remain below 0.2 for the three stars. For the contrast, for the stars HD\,125136, HD\,127195, and HD\,220218 the values are approximately 0.3, 0.1, and 0.45, respectively. The two first values represent weak correlations, with the contrast-RV correlation for HD\,220218 being in the range of moderate correlation. However, a direct interpretation of the value significance remains difficult.

Studying the presence of periodic signals in the RVs using GLS is insightful but not fully conclusive. The Keplerian orbits only become sinusoidal when the eccentricity is zero. The working hypothesis on the noise level, until now assumed as purely photon noise, also shift the FAP threshold levels considerably. Moreover, the absence of correlation between activity indicators and RV is not proof that the signals is not of stellar nature. A full analysis requires the evaluation of a model containing Keplerian signals and noise components. 

\subsection{\texttt{kima} analysis}\label{Sec:kima}

We used the package \texttt{kima}\footnote{\url{https://github.com/j-faria/kima-org/kima}} \citep{2018JOSS....3..487F} to model the planetary orbital parameters, stellar systemic velocity, and instrument-specific noise and RV offset values. In brief, \texttt{kima} uses the diffusive nested sampling algorithm from \cite{2009arXiv0912.2380B} to sample from the parameter posterior distribution; this delivers an estimate for the marginal likelihood for each model and allows for a direct Bayesian model comparison via the marginal likelihoods \citep[see e.g.][]{2014arXiv1411.3921B, 2011MNRAS.415.3462F}. In this context, the number of planets, $N_p$, is a parameter that can be left free or fixed for the analysis. For each model we obtained at least 50\,000 effective samples from the posterior, which allows for a robust sampling of the different parameters.

The classification of a detection with $N_p$ planets as significant depends on the (Keplerian) planet model being favoured relative to the model with $N_{p-1}$ planets, which, in turn, depends on a chosen metric crossing an a-priori threshold. For exoplanet detection within the Bayesian framework, the metric is typically the Bayes factor, and the threshold of 150, following the `very strong evidence' criterion coined by \cite{KassRaftery95}.

We enforced dynamical stability using the angular momentum deficit criterion \citep{2017A&A...605A..72L} by computing the the angular momentum deficit value for all orbital solutions proposed by the sampler and assigning zero likelihood to configurations that violate the stability condition. This provides a simple and computationally efficient way to exclude long-term unstable solutions that would still be compatible with the RV data.

Detailed information on priors and posteriors is provided in Appendix\,\ref{App:kima}. The choice of priors is discussed in Section\,\ref{App:priors} and those adopted are listed in Table\,\ref{Tab:kimapriors}. The posterior distribution median values and credible interval are presented in Table\,\ref{Tab:kimapost} and the corner plots are presented in Appendix\,\ref{App:corners}. 

\setlength{\extrarowheight}{4pt}

\begin{table*}
\centering
\caption{Parameters for the model fitting on the three stars studied in this paper.}
\label{Tab:kimapost}
\begin{tabular}{l c c c c c}
\hline\hline
Parameter$^a$ & Unit & HD\,125136\,b & HD\,127195\,b & HD\,127195\,c & HD\,220218\,b$^b$\ \\[4pt]
\hline

$K$      & [m\,s$^{-1}$]
& $36.28_{-2.26}^{+2.23}$
& $13.89_{-1.83}^{+1.78}$
& $11.56_{-1.82}^{+1.73}$
& $27.48_{-3.77}^{+4.06}$ \\

$P$      & [d]
& $849.77_{-4.73}^{+4.61}$
& $534.50_{-2.92}^{+2.85}$
& $836.99_{-9.67}^{+9.35}$
& $1034.71_{-21.19}^{+17.87}$ \\

$e$      & --
& $0.05_{-0.04}^{+0.06}$
& $0.04_{-0.03}^{+0.05}$
& $0.05_{-0.04}^{+0.05}$
& $0.12_{-0.09}^{+0.12}$ \\

$\phi$   & \multirow{2}{*}{[$^\circ$]}
& $3.18_{-1.54}^{+1.30}$
& $2.66_{-1.75}^{+2.53}$
& $2.83_{-1.39}^{+1.51}$
& $4.35_{-1.29}^{+0.76}$ \\

$\omega$ &
& $3.13_{-2.62}^{+2.67}$
& $3.10_{-1.93}^{+1.97}$
& $4.12_{-2.43}^{+1.21}$
& $1.22_{-0.81}^{+3.98}$ \\[4pt]

\hline

Jitter$_\texttt{COR98}$
& \multirow{7}{*}{[m\,s$^{-1}$]}
& $10.35_{-4.14}^{+13.06}$
& \multicolumn{2}{c}{$13.62_{-6.12}^{+14.51}$}
& -- \\

Jitter$_\texttt{COR07}$
&
& $6.33_{-0.94}^{+1.72}$
& \multicolumn{2}{c}{$7.72_{-1.47}^{+2.04}$}
& $27.57_{-7.04}^{+8.95}$ \\

Jitter$_\texttt{COR14}$
&
& $9.43_{-2.14}^{+2.78}$
& \multicolumn{2}{c}{$8.01_{-1.26}^{+1.52}$}
& $11.23_{-2.27}^{+2.79}$ \\

Jitter$_\texttt{COR24}$
&
& --
& \multicolumn{2}{c}{$11.43_{-3.56}^{+6.13}$}
& $7.83_{-2.00}^{+3.54}$ \\

Offset$_\texttt{COR07-98}$
&
& $5.19_{-16.64}^{+14.66}$
& \multicolumn{2}{c}{$37.26_{-21.42}^{+17.04}$}
& -- \\

Offset$_\texttt{COR14-07}$
&
& $2.95_{-8.47}^{+8.46}$
& \multicolumn{2}{c}{$32.26_{-12.69}^{+10.60}$}
& $4.36_{-25.69}^{+24.61}$ \\

Offset$_\texttt{COR24-14}$
&
& --
& \multicolumn{2}{c}{$18.77_{-9.28}^{+6.31}$}
& $2.87_{-15.88}^{+14.42}$ \\[4pt]

\hline

Slope & \multirow{2}{*}{[m\,s$^{-1}$]} 
      & $0.002_{-0.003}^{+0.003}$ 
      & \multicolumn{2}{c}{$0.003_{-0.002}^{+0.002}$} 
      & $0.000_{-0.005}^{+0.005}$ \\

v$_{\mathrm{sys}}$  &
& $0.68_{-5.36}^{+5.15}$
& \multicolumn{2}{c}{$-21.11_{-7.10}^{+11.27}$}
& $0.40_{-15.78}^{+18.05}$ \\[4pt]

\hline
\end{tabular}
\tablefoot{
\tablefoottext{$a$}{Expected value provided by the median and uncertainties by the 16--84\% credible intervals of the posterior distribution.}
\tablefoottext{$b$}{Activity-rooted signal; see text for details.}
}
\end{table*}

\setlength{\extrarowheight}{2pt}

\begin{figure}[!t]
\center
\includegraphics[width=0.98\columnwidth]{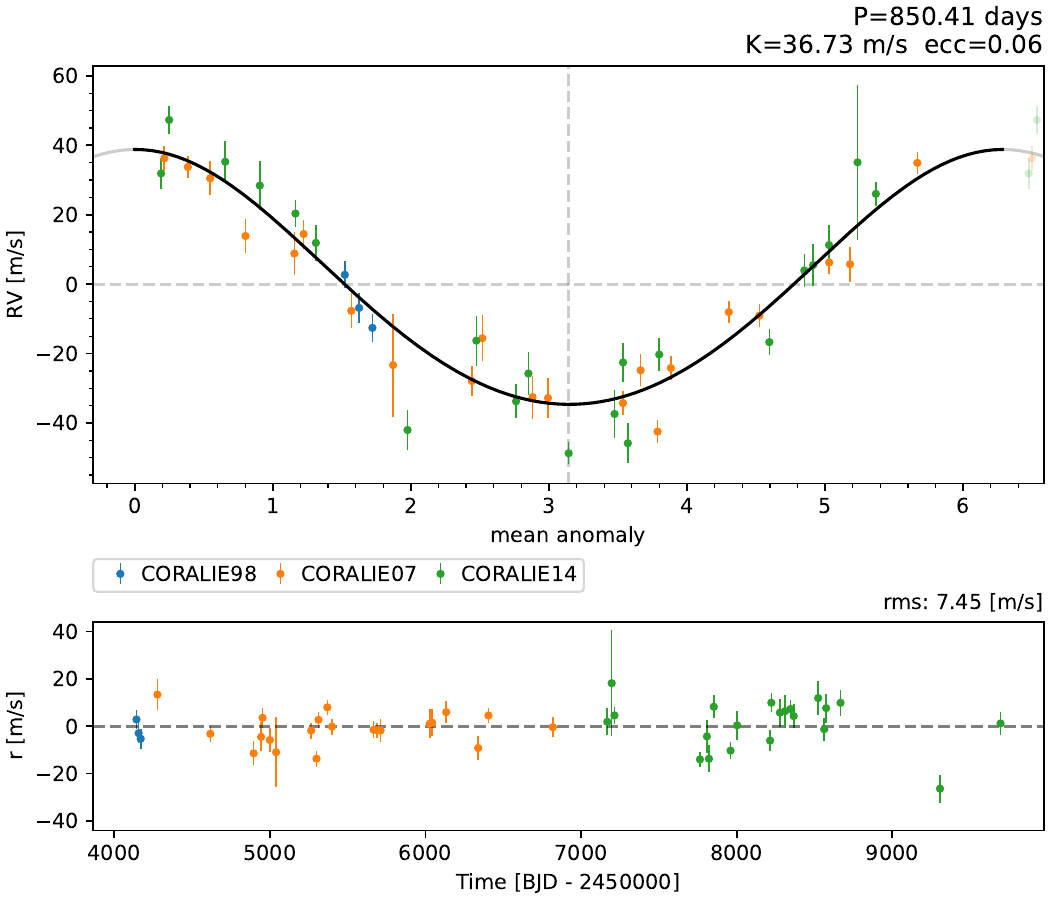}
\caption{Phase plot and residuals as a function of mean anomaly for the planet detected around HD\,125136.}\label{Fig:phaseHD125136}
\centering
\end{figure}

\begin{figure}[!t]
\center
\includegraphics[width=0.98\columnwidth]{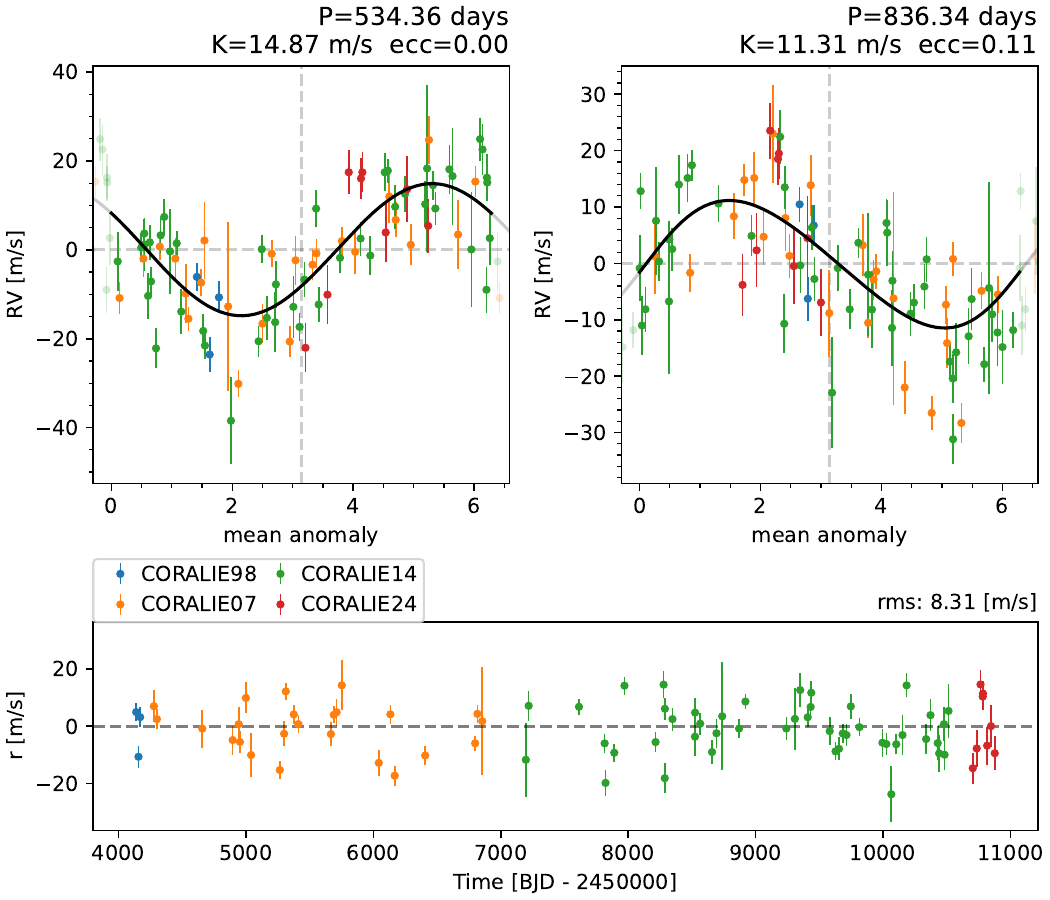}
\caption{Phase plot and residuals as a function of mean anomaly for the two planets detected around HD\,127195.}\label{Fig:phaseHD127195}
\centering
\end{figure}

\begin{figure}[h]
\center
\includegraphics[width=0.98\columnwidth]{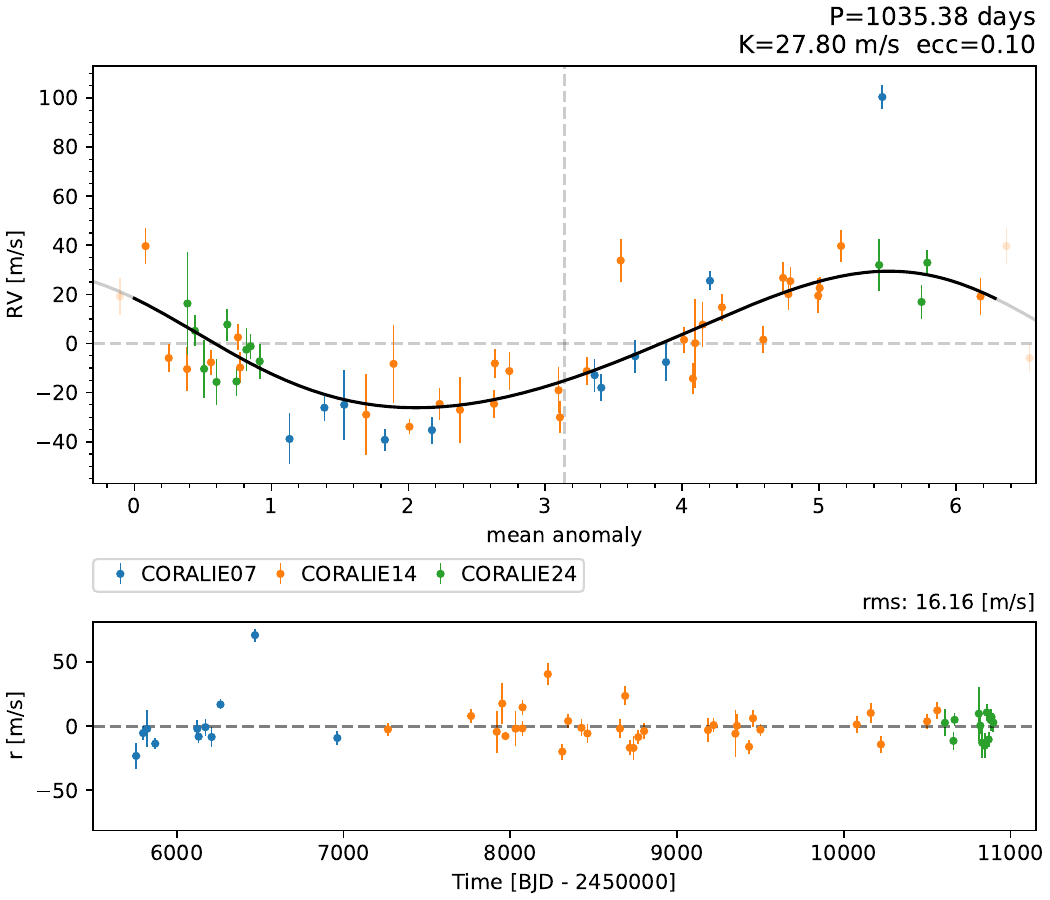}
\caption{Phase plot and residuals as a function of mean anomaly for the signal detected around HD\,220218.}\label{Fig:phaseHD220218}
\centering
\end{figure}

Using a Bayes factor threshold of 150, we confidently detected one signal around the stars HD\,125136 and HD\,220218, and two signals around HD\,127195. The phase plots of the maximum a posteriori (MAP) sample models are shown in Figures\,\ref{Fig:phaseHD125136}, \,\ref{Fig:phaseHD127195}, and \,\ref{Fig:phaseHD220218}. 

It is important to notice that the jitter on HD\,2220218 (7--26\,m\,s$^{-1}$) is noticeably larger than on other stars (6--14\,m\,s$^{-1}$, or 6--11\,{m\,s$^{-1}$ if we exclude the three-point dataset of HD\,127195). The corner plot for HD\,220218 also shows that the amplitude of the RV signal K$_0$ is correlated with other parameters such as $e$ and $\phi$, as well as the jitter term. The fitting of the signal with the current data points and model is thus ambiguous and we did not interpret the signal as being created by a planetary companion. 

\section{Pulsations and achievable RV stability}\label{Sec:PulsPrec}

To estimate the signal introduced by stellar pulsations, we used the model described in \cite{2025A&A...700A.174F}. In a nutshell, the stellar parameters mass, $M$, luminosity, $L$, radius, $R$, and effective temperature, $T_{\mathrm{eff}}$ are used to derive the asteroseismic parameters using robust scaling relations plus reference values calibrated from asteroseismic analyses. These star-tailored values allows us to model the RV signal created by pulsations. The observational campaign on the star is modelled as a sequence of $n$ intra-night integrations with a predefined exposure time, a given time gap between exposures and a photon-noise uncertainty. For a time series acquired in this way, the model predicts the nightly averaged RV (when considering photon noise and pulsations), plus the average intra-night scatter. An installable \texttt{Python} package was made available on \texttt{GitHub}\footnote{\url{https://github.com/pedrorfigueira/rvpulsim}}. 

To test our ability to average out pulsations we selected HD\,127195. This star was chosen from among the three studied in this work because it is slightly brighter than HD\,125136 ($V$\,=\, 7.19\,mag vs 7.44\,mag), with HD\,220218 not being considered for having its RVs dominated by activity.

For HD\,127195 we estimate a pulsation envelope amplitude of 0.9\,m\,s$^{-1}$, populated by 14 modes and a peak period at $\sim$\,1\,h. We tested different observation schemes and selected three observations of 300\,s separated by 1200-1500\,s, assuming a photon noise of 4-4.5\,m\,s$^{-1}$ for each individual exposure. In these conditions, the error on the nightly average introduced by the pulsation is expected to be only $\sim$3\,m\,s$^{-1}$; for comparison, if we used a single integration of 900\,s the value would increase to 5\,m\,s$^{-1}$.

In August 2025, we performed a dedicated observing campaign to emulate this scheme. The star was observed on seven nights over a time span of 23 days; overall the observing conditions were good (transparency only reduced by thin clouds, seeing $<$1\arcsec), and the target precision was attained, albeit with some variance. The nightly scatter ranged from 2 to 8 m\,s$^{-1}$, very probably due to variable observing conditions, but the median value was 4\,m\,s$^{-1}$, the expected value from simulations.

To test whether this observing strategy was successful at reducing the RV scatter, we repeated the analysis performed in Sect.\,\ref{Sec:kima}, now including the dataset \texttt{CORALIE25} composed of the nightly averaged observations carried out with the new strategy. The results are presented in Table\,\ref{Tab:puls}. The planetary parameters and instrument characterisation for the datasets \texttt{CORALIE98} to \texttt{CORALIE24} are, within their credible interval, fully compatible with those derived previously. The MAP solution provides both a jitter and an rms of the residuals of 5\,m\,s$^{-1}$, significantly lower than for the other datasets (see Table\,\ref{Tab:rms_jitter_2025}, and Fig.\,\ref{Fig:phaseHD127195_Camp25}). This aligns with our predictions, with a stellar contribution to the uncertainties of 3\,m\,s$^{-1}$ and an instrumentation stability contribution of 4\,m\,s$^{-1}$. Conversely, if the instrument stability is of only 3\,m\,s$^{-1}$, this means our stellar signal is still at 4\,m\,s$^{-1}$. 

To our knowledge, there are only a few cases in which an RV precision below 5\,m\,s$^{-1}$ has been achieved for giant stars. A recent example is the work of \cite{2024AJ....168....1G} on TIC\,365102760, which, through a joint analysis of photometry and RV data, reports a jitter of 4.2\,m\,s$^{-1}$.

\section{Discussion}\label{Sec:Disc}

\subsection{Detected signals: Planets or stellar activity}

Stellar phenomena can create false positive signals and some can reach timescales of hundreds of days, as we modelled in this work. Radial pulsations are directly excluded from the simulations in Sect.\,\ref{Sec:PulsPrec} and works such as \cite{1980tsp..book.....C}, which show that pulsation timescales do not exceed $\sim$1\,day. Low-luminosity red giant stars have measured surface rotation periods primarily in the range of $\sim$10–200\,d, based on spot-modulation analyses of \textit{Kepler} light curves \citep[e.g.][]{2017A&A...605A.111C, 2020A&A...639A..63G}, with a detection tail extending toward longer periods but strongly affected by incompleteness. Since only a small fraction of red giant stars exhibit detectable photometric modulation, likely due to weak magnetic activity and finite spot lifetimes, rotation periods longer than $\sim$200\,d may remain undetected in current photometric data. A realistic rotational period estimate can be obtained from stellar evolution models including angular momentum transport in radiative interiors \citep[e.g.][]{2012A&A...539A..70E, 2013ApJ...775L...1T}. These models include enhanced angular momentum transport that leads to partial core--envelope coupling, as required by asteroseismic constraints. While a simple scaling based on envelope expansion ($P \propto R^2$) yields an upper bound for rotation periods on the red giant branch $P_{\rm RGB} \sim 300$--500\,d when considering a star evolving from the main sequence and associated period $P_{\rm MS} \sim 20$--30\,d and reaching $R \sim 4\,R_\odot$, models including angular momentum transport typically predict surface rotation periods of $\sim30$--150\,d at this evolutionary stage. This range is consistent with forward modelling predictions and suggests that very long periods ($\gtrsim 300$\,d) are likely associated with weak internal coupling or unusually low initial angular momentum.

The other candidate mechanism is magnetic cycles. \cite{2011arXiv1107.5325L} studied the chromospheric activity in $\sim$300 stars over 2--7\,yr and concluded that while 61$\pm$8\% of FGK main sequence stars show a magnetic cycle, this number drops to 12$\pm$5\% for subgiants. Their results are in line with previous studies that found that stars show lower chromospheric activity as they evolve off the main sequence \citep[e.g.][]{2004AJ....128.1273W}. The work of \cite{2021A&A...646A..77G} confirmed this result with a much larger sample, and even showed that the few evolved stars with high chromospheric activity were unusually massive or young, as proposed by \cite{2020AJ....159..235L}.
\cite{2011arXiv1107.5325L} also found that the sensitivity of RV to cycles is almost zero for T$_{\mathrm{eff}}\,\sim$\,4800\,K, a value very similar to that of our stars. Magnetic cycles are less likely to have a significant impact on RV if present. They cannot be discarded as potential mechanisms generating the measured signals, given that their typical timescales overlap with the observed periodicities.

There is an alternative mechanism worth mentioning. \cite{2024A&A...689A..91S} explored the origin of a periodic signal with 674\,d measured on the giant star NGC 4349 No. 127 \citep{2018PASP..130i4202D, 2023A&A...679A..94D}. The authors showed that the RV and activity-indicator time series, as well as the correlations between them, were aptly reproduced by non-radial pulsations. Depending on the angular degree, $l$, and azimuthal mode, $m$, the correlation with indicators can show very different strengths, being intrinsically low in some cases ($m$\,=\,0). The authors were motivated by the work of \cite{2015MNRAS.452.3863S}, who proposed oscillatory convective modes as the mechanism explaining long secondary periods of 200--1500\,d measured on the photometry of low-mass red giants. 
If we extrapolate the period estimations of \cite{2015MNRAS.452.3863S} to our luminosity domain (or use the empirical rule that long secondary periods have periods of 5--10 times the fundamental radial pulsation), we would obtain periods of the order of a few tens of days; thus, this is unlikely to be the explanation behind the observed signals at $P$\,$>$500\,d.

Considering the different scenarios, we therefore conclude the signals observed in HD\,125136 and HD\,127195 are very probably of planetary origin. Our only reservation is that their periods fall within the range of long-term activity cycles, which are nonetheless know to have weak RV signatures on subgiant stars. The planet around HD\,125136 has a minimum mass $m\sin{i}$ of 2.26$\pm$0.08\,$M_\mathrm{Jup}$ and a period of 850\,d; for HD\,127195, the planets have minimum masses of 0.67$\pm$0.02\,$M_\mathrm{Jup}$ and 0.65$\pm$0.02\,$M_\mathrm{Jup}$ and with orbital periods of 535\,d and 837\,d, respectively. The signal detected on HD\,220218 shows correlations between the semi-amplitude and other orbital parameters. It can be fitted with a jitter much larger than is customary on giant and subgiant stars observed with CORALIE. We consider this a spurious fit, likely driven by stellar activity, and do not discuss it further in this work. 

\subsection{Planetary orbits through stellar evolutionary phases}\label{Sec:planevol}

To explore the transformation of planetary orbits during stellar evolution, we used the framework presented in \cite{2026NatAs..10..124E}. HD\,125136 and HD\,127195 were simulated as stars with an initial mass of 1.5\,$M_\odot$, and the current time was estimated as the time where the current radius of the star matches the radius of the stellar evolution model. The resulting orbital evolution for HD\,125136\,b, HD\,127195\,b and c is shown in Fig.~\ref{Fig:orbitalevolution}.

The planet HD\,125136\,b will survive the red giant branch (RGB) phase and due to the mass-loss the planetary orbit will widen during this phase. However, during the asymptotic giant branch (AGB) phase, tides will take over and the planet will spiral inward and be consumed by the star. It will thus not survive until the white dwarf (WD) phase. The system around HD\,127195 will undergo similar evolutionary effects over the same milestones but with a different final outcome. As the star enters the RGB phase, HD\,127195\,b will experience a slight increase in its semi-major axis due to stellar mass-loss, but will return close to its initial value as tides become important. The duration of the RGB phase is sufficiently short for the planet to survive through it. During the AGB phase however, tides continue to shrink the semi-major axis until it is swallowed by the star, similar to HD\,125136\,b. On the other hand, as planet c starts on a similar orbit as HD\,125136\,b, their evolution until the AGB phase will be similar. However, as HD\,127195\,c has a lower mass, tides will have a weaker impact, allowing mass-loss to widen the orbit and permitting the survival of the planet up to the WD phase. 

\begin{figure}[h]
\center
\includegraphics[width=0.8\columnwidth]{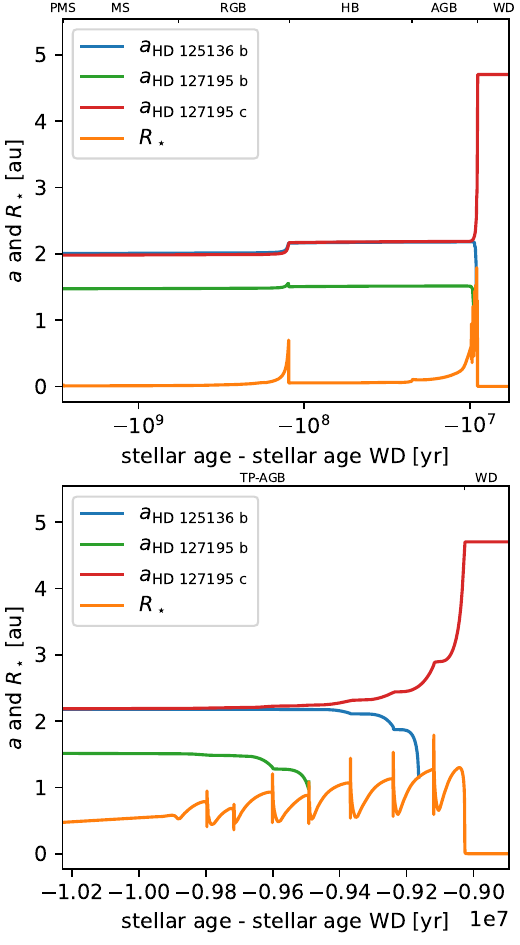}
\caption{Semi-major axis of planetary orbit and stellar radius as a function of stellar age for the systems HD\,125136 and HD\,127195. The orbits of the three planets are presented in different colours. We note that the $R_*$ value is the same for both stars.}\label{Fig:orbitalevolution}
\centering
\end{figure}

\subsection{RV noise floor for evolved stars and achievable precision}

The rms of the residuals on the planetary models is 7--8\,m\,s$^{-1}$, in excess of the photon noise of $\sim$5\,m\,s$^{-1}$. The extra component on the error budget might be due to the instrument or intrinsic stellar RV stability, with the most likely contributor being stellar p-mode oscillations. 

\cite{2018MNRAS.480L..48Y} estimated the impact of oscillations on RV of dwarf to giant stars using \textit{Kepler} data and re-fitting scaling relationships. The authors estimated an RV scatter of 1.5\,m\,s$^{-1}$ on subgiants and of 4\,m\,s$^{-1}$ for low-luminosity giants. These values are lower than those measured on subgiant surveys, as presented by \cite{2010PASP..122..701J} and  \cite{2011ApJS..197...26J}, that estimated the noise floor to be at 5\,m\,s$^{-1}$ when accounting for the various sources of stellar variability.

\cite{2020AJ....159..235L} measured a RV jitter of 5--10\,m\,s$^{-1}$ for slightly evolved stars (subgiants and low-luminosity giants) with masses similar to those of our targets. The authors also predict the dependence on stellar parameters using a chromospheric term and dividing the convective component into oscillation and granulation. \cite{2025A&A...700A.174F} showed that the granulation signature of the main sequence chromospherically quiet $\tau$ Cet (K0V) is lower than predicted from the Sun. We can speculate that chromospherically quiet K evolved stars also exhibit a reduced granulation component and, thus, pulsations become the main engine behind RV variability. 

Theoretical works like that \cite{1997A&A...317..723T} and observational campaigns such as the one reported by \cite{2025ApJ...987..168L} on the subgiant HD\,142091 showed that pulsation signals present negligible line profile variations and very weak chromatic dependence, making them difficult to distinguish once detected. 
The results from the observing campaign described in Sect.\,\ref{Sec:PulsPrec} show this effect can be mitigated efficiently by the observing strategy. The 4\,m\,s$^{-1}$ scatter achieved in this first test encourages applying this strategy at higher photon-noise precision and over longer observing baselines.

\subsection{Compatibility limits and detectable planets}\label{Sec:compatlim}

When running \texttt{kima}, it is possible to fix the parameter $N_p$ to the number of detected planets plus one, so that we can estimate the parameters of undetected planets that are compatible with the data \citep[see][]{2022MNRAS.511.3571S}. The samples with the largest semi-amplitude $K_0$ for a given period bin correspond to the most massive planet compatible with the data and the adopted priors. The compatibility limits produced in this way are, by construction, marginalised over all parameters other than mass and period. In these simulations, we set as the stellar masses the values presented in \cite{2022A&A...657A..87O}, listed in Table\,\ref{Tab:SteParams}.  We present the compatibility limits in Fig.\,\ref{Fig:CompatLimits}.

\begin{figure}[h]
\includegraphics[width=8.2cm]{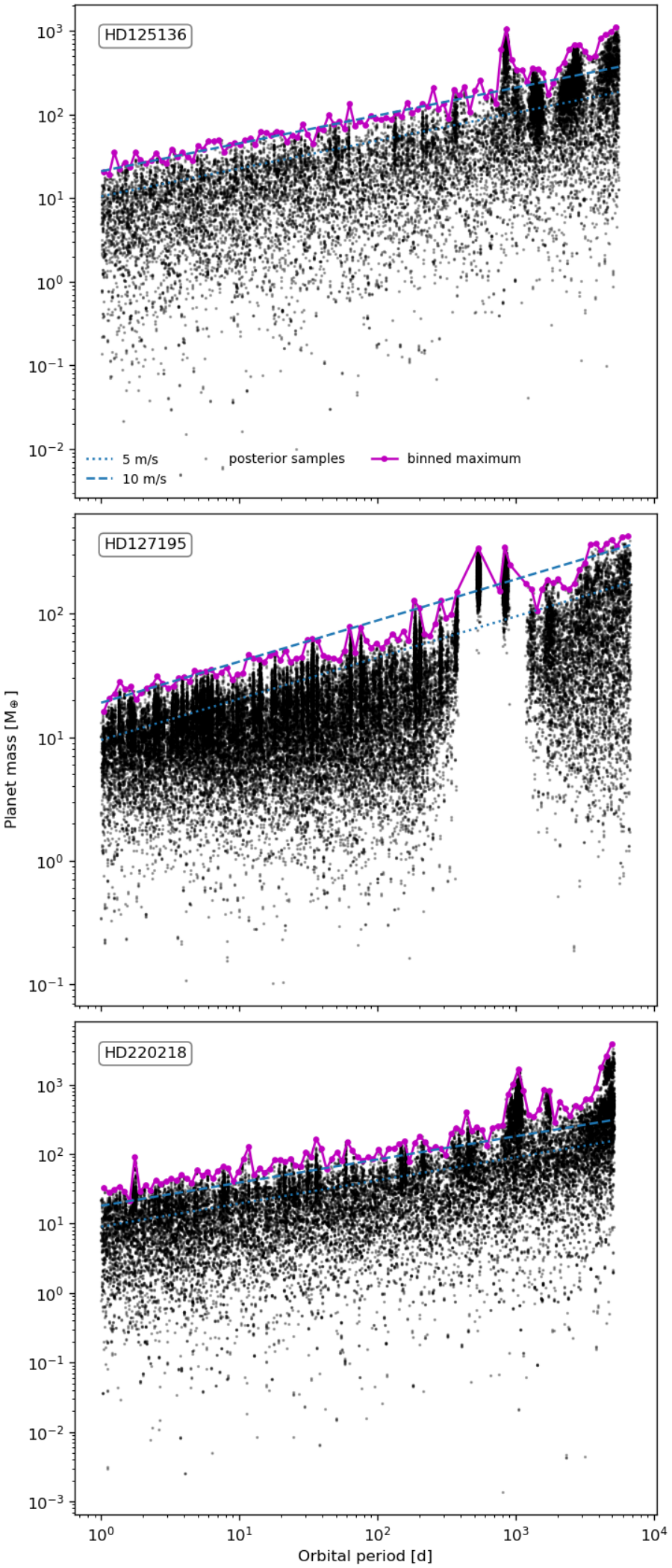}
\caption{Compatibility limits for the three stars studied, HD\,125136 (\textit{top}), HD\,127195 (\textit{middle}), and HD\,220218 (\textit{bottom}). The maximum RV amplitude of the samples is plotted along with lines corresponding to a mass with amplitude equal to a precision limit of 5 and 10\,m\,s$^{-1}$.}\label{Fig:CompatLimits}
\centering
\end{figure}

The compatibility limits show that we are confidently able to detect planets with an amplitude larger than 10\,m\,s$^{-1}$. This is in excess of the $\sim$5\,m\,s$^{-1}$ of precision of CORALIE and photon-noise uncertainty of our measurements, but in line with the expectations that subgiant stars have an additional source of noise up to $\sim$10\,m\,s$^{-1}$ (see previous section). We were able to detect planets with roughly twice the mass of Neptune on short orbits of $\sim$10\,d and Saturn-mass planets up to 100\,d. For some stars, we were able to detect Jupiter-mass planets with orbit periods up to $\sim$1000\,d, but for others, we found that only planets with more than twice the mass of Jupiter are detectable. The resulting values will depend on the characteristics of star and of the observations, but the cases studied here provide representative order-of-magnitude estimates.

\subsection{Demographics of planets around evolved stars}

Dedicated RV surveys have measured the occurrence rate of planets around evolved stars \citep{2015A&A...574A.116R, 2018ApJ...860..109G}, with \cite{2022A&A...661A..63W} pooling together different surveys. Today we know that giant-planet occurrence rises with stellar mass from $\leq$1\,$M_{\odot}$ up to 1.7–1.9\,$M_{\odot}$ (with a peak occurrence of 10–20\% in this mass domain), decreasing very sharply for higher mass hosts. The difference in occurrence rates reported by different surveys can be explained by different choices in sample selection and completeness corrections. \cite{2022A&A...661A..63W} performed the most homogeneous analysis and reaches an overall giant planet occurrence of 10.7\% that peaks at 1.68$\pm$0.59\,$M_{\odot}$. Interestingly, the authors also found that the period distribution for detected giants peaks at 700–800\,d where the occurrence is higher than reported for main sequence FGK samples. This result had also been presented by \cite{2015A&A...574A.116R} for comparable periods and masses. Although this higher occurrence is expected to be real (the authors carefully accounted for observational biases), there is always the possibility of false positives, as shown recently by \cite{2025A&A...699A.260H}.   

In Fig.\,\ref{Fig:planetsGiants}, we represent the planets currently know to orbit stars with a mass above 1.1\,$M_{\odot}$. This corresponds to $\sim$21\% of the planets known, with the percentage being reduced to $\sim$11\% when restricting the sample to planets with masses measured via RV. The planets presented in this work fall close to this occurrence 700--800\,d peak in period space, yet they are in an area that remains sparsely populated. The diagram is a stark reminder of how hard it is to follow stars for timescales of decades with stable RV spectrographs. While we know that giant planets have higher occurrence rates around giant stars and sub-Neptunes around M-dwarfs \citep[see e.g. review by][]{2018haex.bookE.153M} and that giant planet occurrence peaks near the snow line \citep[e.g.][]{2019ApJ...874...81F}, it remains unclear whether the apparent dearth of Neptune-mass planets at comparable separations is a consequence of detection bias or a direct outcome of planet formation.

\begin{figure*}
\center
\includegraphics[width=0.8\textwidth]{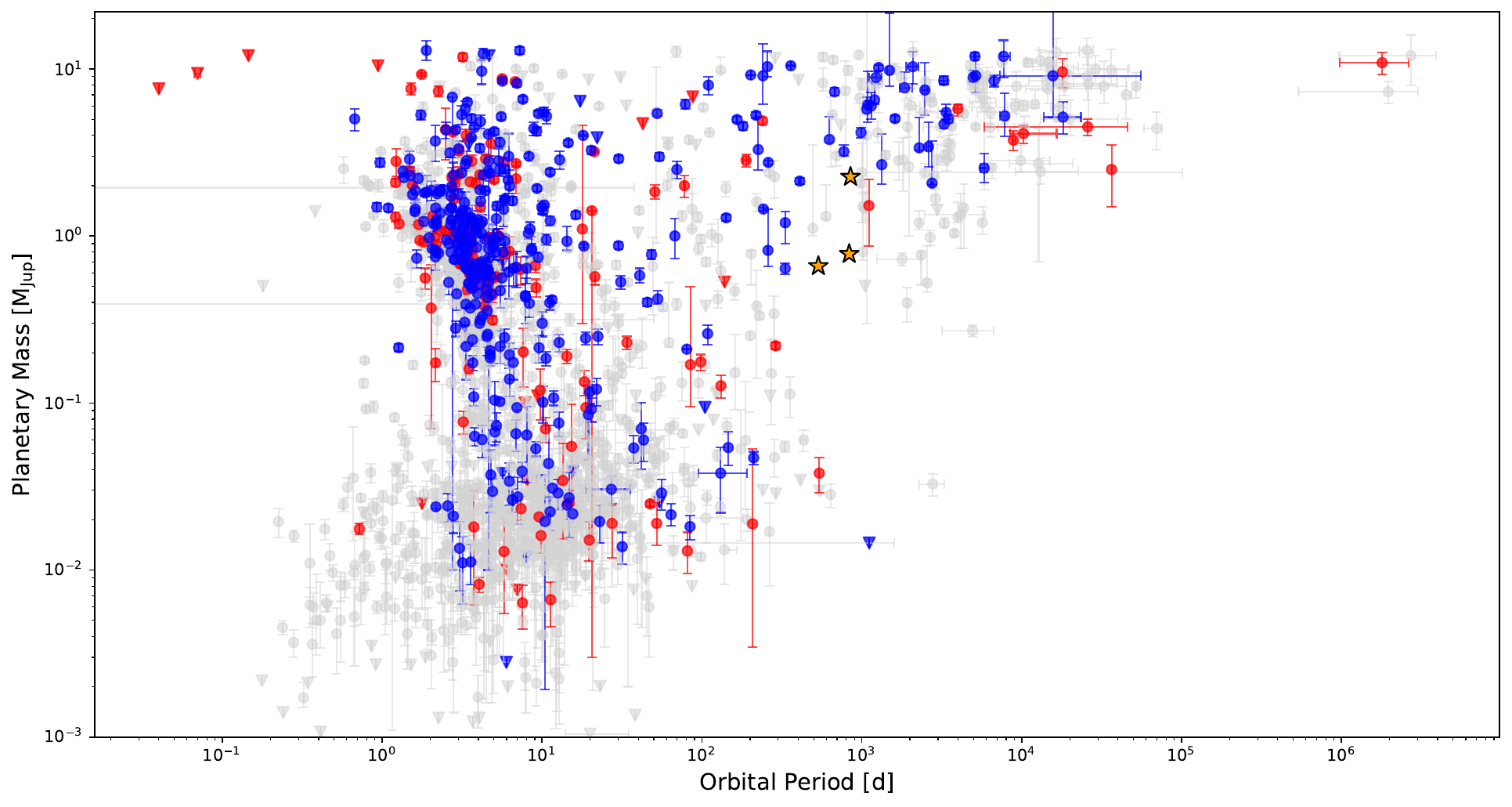}
\caption{Mass-period diagram for RV-characterised exoplanets (blue) or with other methods (red) around hosts with mass $M_*$\,>\,1.1\,$M_{\odot}$, when compared with other known planets (grey). Only objects with planetary mass $m$\,<\,13\,$M_{\mathrm{Jup}}$ are considered; mass upper limits are marked with inverted triangles. The three planets presented in this paper are represented as orange stars.}\label{Fig:planetsGiants}
\centering
\end{figure*}

Transit surveys, which are primarily sensitive to short-period systems, provide very different yet complementary constraints on the occurrence of planets around evolved stars. The studies of \cite{2019AJ....158..227G} and \cite{2023A&A...670A..26T} used \textit{Kepler} and TESS data, respectively, to evaluate the occurrence rate of short-period giant planets and compare the results with the values obtained around main sequence stars. The occurrence rates remained compatible, suggesting that the effects of stellar evolution on the tidal decay and engulfment of Jupiter-mass close-in (P\,$<$\,10\,d) planets are not significant until stars leave the low-luminosity giant branch regime. However, the more recent work of \cite{2025MNRAS.544.1186B} characterised a larger number of stars and determined more precise occurrence rates. The study measured a significant decrease between subgiants and low-luminosity giants, along with a decrease when subgiants were compared with main sequence planets, with a higher occurrence rate decrease measured for shorter period planets. This provided clear evidence that the population of short-period giant planets is being sculpted by the post-main sequence evolution of the host stars, as discussed for our longer period planets in Section\,\ref{Sec:planevol}.

\section{Conclusions}\label{Sec:Conc}

We analysed data acquired with the CORALIE spectrograph in the context of the survey CASCADES on three subgiant stars: HD\,125136, HD\,127195, and HD\,220218. The RV time series, spanning 15--18\,yr, revealed the presence of several stellar signals. We interpret the signal on HD\,125136 as arising from a planet with a minimum mass of 2.26$\pm$0.08\,$M_\mathrm{Jup}$ and a period of 850\,d. On HD\,127195 we detect a two-planet system, with the planets having a minimum mass of 0.67$\pm$0.02\,$M_\mathrm{Jup}$ and 0.65$\pm$0.02\,$M_\mathrm{Jup}$ with orbital periods of 535\,d and 837\,d, respectively. There is no correlation between RV and activity indicators time series and the latter do not show any significant periodicity. However, we cannot fully exclude the possibility that these signals are false positives and, in particular, generated by long-term magnetic cycles. The signal around HD\,220218 is interpreted as spurious and very likely caused by stellar activity.

The three planets around HD\,125136 and HD\,127195 fall on a parameter space poorly populated in stellar mass and period. The systems reported here therefore provide useful constraints on the population of Saturn and Jupiter planets around evolved stars. We showed that on the subgiant HD\,127195 a dedicated observing strategy can efficiently average out  stellar pulsations to below 5\,m\,s$^{-1}$. Such strategies will be increasingly important for extending RV planet searches to evolved stars with a higher intrinsic variability.

\section*{Data availability}

The RV data for the targets discussed in this work are available in electronic form at the CDS via anonymous \texttt{ftp} to \url{cdsarc.u-strasbg.fr} (130.79.128.5) or via \url{http://cdsweb.u-strasbg.fr/cgi-bin/qcat?J/A+A/}.

\begin{acknowledgements}

    We thank the anonymous referee for insightful and very valuable comments on the first version of the manuscript.
    
    Based on observations collected with CORALIE echelle spectrograph mounted on the Euler 1.2\,m Swiss telescope at La Silla, Chile.

    PF thanks P. Eggenberger for very interesting discussions on stellar pulsations.

    PF acknowledges support from the COST Action CA22133 PLANETS. PF and RL acknowledges financial support from the Severo Ochoa grant CEX2021-001131-S funded by MCIN/AEI/10.13039/501100011033. PF is also funded by the European Union (ERC, THIRSTEE, 101164189). Views and opinions expressed are however those of the author(s) only and do not necessarily reflect those of the European Union or the European Research Council. Neither the European Union nor the granting authority can be held responsible for them.

    ME acknowledges support from the FWO grants G0B3823N and G099720N.
    
    MS thanks the Belgian Federal Science Policy Office (BELSPO) for the provision of financial support in the framework of the PRODEX Programme of the European Space Agency (ESA) under contract number C4000140754. 

    This work has made use of data from the European Space Agency (ESA) mission {\it Gaia} (\url{https://www.cosmos.esa.int/gaia}), processed by the {\it Gaia} Data Processing and Analysis Consortium (DPAC, \url{https://www.cosmos.esa.int/web/gaia/dpac/consortium}). Funding for the DPAC has been provided by national institutions, in particular the institutions participating in the {\it Gaia} Multilateral Agreement.
        
    This work made use of {\tt astropy}, a community-developed core Python package and an ecosystem of tools and resources for astronomy, described in  \href{https://ui.adsabs.harvard.edu/abs/2013A%26A...558A..33A/abstract}{astropy collaboration (2013)}, \href{https://ui.adsabs.harvard.edu/abs/2018AJ....156..123A/abstract}{astropy collaboration (2018)}, and  \href{https://ui.adsabs.harvard.edu/abs/2022ApJ...935..167A/abstract}{astropy collaboration (2022)}. 
    
    Corner plots were created using the package \texttt{corner.py}, described in \href{https://doi.org/10.21105/joss.00024}{Foreman-Mackey (2016)}.
            
\end{acknowledgements}

\bibliographystyle{aa}
\bibliography{bibliography.bib}

\begin{appendix} 
\nolinenumbers

\section{Stellar characterisation}\label{App:SteChar}

\subsection{\textit{Gaia} DR3 data}

We used the \texttt{astroquery}\footnote{\url{https://astroquery.readthedocs.io/}} package \citep{2019AJ....157...98G} to download the \textit{Gaia} data on our targets, complementing it with targets of the solar neighbourhood for illustrative purposes (Fig\,\ref{Fig:CASCADES_vs_Gaia}).
On top of the \texttt{GPS-PHOT} model that delivers parameters from SED fitting, we select stellar Luminosity, mass, radius and $\log{g}$ from the \texttt{FLAME} package, designed to provide reliable estimates for evolved stars. These use evolutionary models, capturing post-MS changes to avoid photometric degeneracies and and ensure consistent modelling \citep{2023A&A...674A..26C}. 

\subsection{\texttt{ARIADNE} parameters derivation}

We used the \texttt{ARIADNE}\footnote{\url{https://github.com/jvines/astroARIADNE}} package described in \cite{2022MNRAS.513.2719V} to fetch photometric data of our targets from public surveys and perform spectral energy distribution (SED) fitting of stellar models. The photometry for the three stars is presented on Table\,\ref{Tab:ARIADNEinput}.

\begin{table}[h]
\centering
\caption{\texttt{ARIADNE} input photometry fetched for HD125136, HD127195, and HD220218$^\dagger$.}
\label{Tab:ARIADNEinput}
\resizebox{\columnwidth}{!}{%
 \begin{tabular}{l ccc}
 \hline  \hline
Filter      & HD125136 & HD127195 & HD220218 \\ \hline
GALEX NUV        & 15.2180$\pm$0.0090 & $\ldots$ & 15.4240$\pm$0.0140 \\
Strömgren $u$    & 10.4260$\pm$0.0000 & $\ldots$ & 10.8670$\pm$0.0000 \\
Strömgren $v$    & 9.0330$\pm$0.0000 & $\ldots$ & 9.5560$\pm$0.0000 \\
Tycho $B_T$      & 8.6950$\pm$0.0160 & 8.4040$\pm$0.0160 & 9.2200$\pm$0.0170 \\
Johnson $B$      & 8.4180$\pm$0.0100 & 8.1330$\pm$0.0100 & 8.9840$\pm$0.0140 \\
Strömgren $b$    & 8.0450$\pm$0.0000 & $\ldots$ & 8.6640$\pm$0.0000 \\
\textit{Gaia} DR3 $G_{\rm BP}$ & 7.6689$\pm$0.0028 & 7.4116$\pm$0.0028 & 8.3037$\pm$0.0028 \\
Tycho $V_T$      & 7.5490$\pm$0.0100 & 7.3000$\pm$0.0100 & 8.1660$\pm$0.0110 \\
Strömgren $y$    & 7.4400$\pm$0.0000 & $\ldots$ & 8.0840$\pm$0.0000 \\
Johnson $V$      & 7.4330$\pm$0.0080 & 7.1840$\pm$0.0070 & 8.0620$\pm$0.0080 \\
\textit{Gaia} DR3 $G$     & 7.1869$\pm$0.0028 & 6.9385$\pm$0.0028 & 7.8373$\pm$0.0028 \\
TESS             & 6.5875$\pm$0.0060 & 6.3478$\pm$0.0060 & 7.2539$\pm$0.0060 \\
\textit{Gaia} DR3 $G_{\rm RP}$ & 6.5385$\pm$0.0038 & 6.2965$\pm$0.0038 & 7.2004$\pm$0.0038 \\
2MASS $J$        & 5.7610$\pm$0.0190 & $\ldots$ & 6.4380$\pm$0.0240 \\
2MASS $H$        & 5.2700$\pm$0.0440 & 5.0840$\pm$0.0180 & 5.9450$\pm$0.0260 \\
2MASS $K_{\rm s}$ & 5.1440$\pm$0.0200 & 4.9140$\pm$0.0210 & 5.8320$\pm$0.0200 \\

\hline
\end{tabular}}
\tablefoot{
\tablefoottext{$\dagger$}{The \textit{Gaia} DR3 ids for HD\,125136, HD\,127195, HD\,220218 are 5797485811232714112, 6091942515572349312, and 6490287572484640896, respectively.}
}
\end{table}

The programme fits the SED across the fetched photometric bands and uncertainties using several stellar atmosphere models and employing Bayesian Model Averaging (BMA). For each star, the final parameter posterior is created by drawing from the posterior of the different models, weighted by their evidence. Model averaging is done over PHOENIX \citep{2013A&A...553A...6H}, BT-Settl \citep{2012RSPTA.370.2765A}, ATLAS9 \citep{2003IAUS..210P.A20C}, and SYNTHE Kurucz \citep{1993sssp.book.....K} models. \texttt{ARIADNE} then queries \textit{Gaia} DR3 for parameters, and provides the best estimate for several other stellar parameters. 
We use the default (weakly informative) priors on the parameters to fit. Extinction is fitted using the dustmap \texttt{SFD} from the package \texttt{dustmaps}\footnote{\url{https://dustmaps.readthedocs.io/en/latest/maps.htm}}.

\subsection{Comparison between CASCADES, \textit{Gaia} DR3, and \texttt{ARIADNE} data} 

The best-fit stellar parameters $T_{\mathrm{eff}}$. $\log{g}$, [Fe/H], distance d, luminosity $L$, radius $R$, mass $M$ and Age are presented on Table\,\ref{Tab:SteParams} and represented graphically in Fig\,\ref{Fig:SteParams}.  

\texttt{ARIADNE} seems to overestimate $\log{g}$ when compared with the other two datasets. The $T_{\mathrm{eff}}$, luminosity and mass for HD\,220218 from \textit{Gaia} DR3 are significantly larger than those derived with the other methods. Finally, the \textit{Gaia} formal uncertainties on $T_{\mathrm{eff}}$ and $\log{g}$ are much smaller than 1\%, and as a consequence we are certainly limited by the method's accuracy, as discussed in \cite{2022ApJ...927...31T}. 

\setlength{\extrarowheight}{3pt}

\begin{table*}[h]
\centering
\caption{Comparison of stellar parameters from CASCADES, \textit{Gaia}, and \texttt{ARIADNE}.}
\label{Tab:SteParams}
\begin{tabular}{llccc}
\hline\hline
Parameter & Method & HD\,125136 & HD\,127195 & HD\,220218 \\
\hline
\multirow{3}{*}{$T_{\mathrm{eff}}$ \small{[K]}} & CASCADES & $4959 \pm 34$ & $4981 \pm 26$ & $4962 \pm 27$ \\
 & \textit{Gaia} & $4968.4_{-1.0}^{+1.2}$ & $5062.2_{-1.5}^{+1.7}$ & $5174.3_{-2.9}^{+3.2}$ \\
& \texttt{ARIADNE} & 4928.20$^{+53.34}_{-45.87}$ & 5080.98$^{+119.04}_{-111.83}$ & 4966.95$^{+41.22}_{-41.22}$ \\[5pt]

\multirow{2}{*}{[Fe/H]} & CASCADES & $0.02 \pm 0.03$ & $-0.04 \pm 0.02$ & $-0.17 \pm 0.02$ \\
& \texttt{ARIADNE} & 0.08$^{+0.13}_{-0.10}$ & -0.13$^{+0.17}_{-0.17}$ & -0.08$^{+0.10}_{-0.11}$ \\[5pt]

\multirow{3}{*}{$\log{g}$} & CASCADES & $3.11 \pm 0.07$ & $3.29 \pm 0.08$ & $3.16 \pm 0.08$ \\
& \textit{Gaia} & $3.1090_{-0.0015}^{+0.0016}$ & $3.3279_{-0.0037}^{+0.0026}$ & $3.2567_{-0.0027}^{+0.0025}$ \\
& \texttt{ARIADNE} & 4.07$^{+0.41}_{-0.34}$ & 4.26$^{+0.46}_{-0.39}$ & 4.16$^{+0.43}_{-0.40}$ \\[5pt]

\multirow{3}{*}{d \small{[pc]}} & CASCADES & $121.3 \pm 0.4$ & $84.7 \pm 0.6$ & $147.6 \pm 0.7$ \\
 & \textit{Gaia} & $119.97_{-0.20}^{+0.20}$ & $83.93_{-0.23}^{+0.35}$ & $145.09_{-0.42}^{+0.45}$ \\
 & \texttt{ARIADNE} & 121.21$^{+0.34}_{-0.20}$ & 84.43$^{+0.37}_{-0.29}$ & 148.76$^{+0.94}_{-0.55}$ \\[5pt]

\multirow{3}{*}{$L$ \small{[$L_{\odot}$]}} & CASCADES & $15.80 \pm 0.28$ & $9.61 \pm 0.19$ & $13.18 \pm 0.24$ \\
 & \textit{Gaia} & $17.46_{-0.06}^{+0.06}$ & $10.92_{-0.05}^{+0.05}$ & $15.23_{-0.09}^{+0.09}$ \\
& \texttt{ARIADNE} & 16.89$^{+1.00}_{-0.88}$ & 10.65$^{+1.25}_{-1.05}$ & 13.63$^{+0.71}_{-0.76}$ \\[5pt]

\multirow{3}{*}{$M$ \small{[$M_{\odot}$]}} & CASCADES & $1.55 \pm 0.08$ & $1.34 \pm 0.06$ & $1.24 \pm 0.05$ \\
 & \textit{Gaia} & $\ldots$ & $1.60_{-0.04}^{+0.04}$ & $1.85_{-0.05}^{+0.04}$ \\
 & \texttt{ARIADNE} & 1.27$^{+0.35}_{-0.24}$ & 1.39$^{+0.28}_{-0.32}$ & 1.24$^{+0.27}_{-0.22}$ \\[5pt]

\multirow{3}{*}{$R$ \small{[$R_{\odot}$]}} & CASCADES & $5.39 \pm 0.09$ & $4.16 \pm 0.06$ & $4.91 \pm 0.07$ \\
 & \textit{Gaia} & $5.64_{-0.11}^{+0.11}$ & $4.30_{-0.09}^{+0.09}$ & $4.86_{-0.10}^{+0.10}$ \\
 & \texttt{ARIADNE} & 5.64$^{+0.10}_{-0.11}$ & 4.21$^{+0.14}_{-0.11}$ & 4.98$^{+0.11}_{-0.10}$ \\[5pt]

\multirow{2}{*}{Age \small[Gyr]} & \textit{Gaia} & $\ldots$ & $2.3_{-0.3}^{+0.2}$ & $1.5_{-0.2}^{+0.2}$ \\
 & \texttt{ARIADNE} & 4.03$^{+2.36}_{-2.65}$ & 2.34$^{+2.61}_{-0.978}$ & 3.99$^{+2.69}_{-2.02}$ \\[4pt]

\hline
\end{tabular}
\end{table*}

\setlength{\extrarowheight}{2pt}

\begin{figure*}[h]
\center
\includegraphics[width=\textwidth]{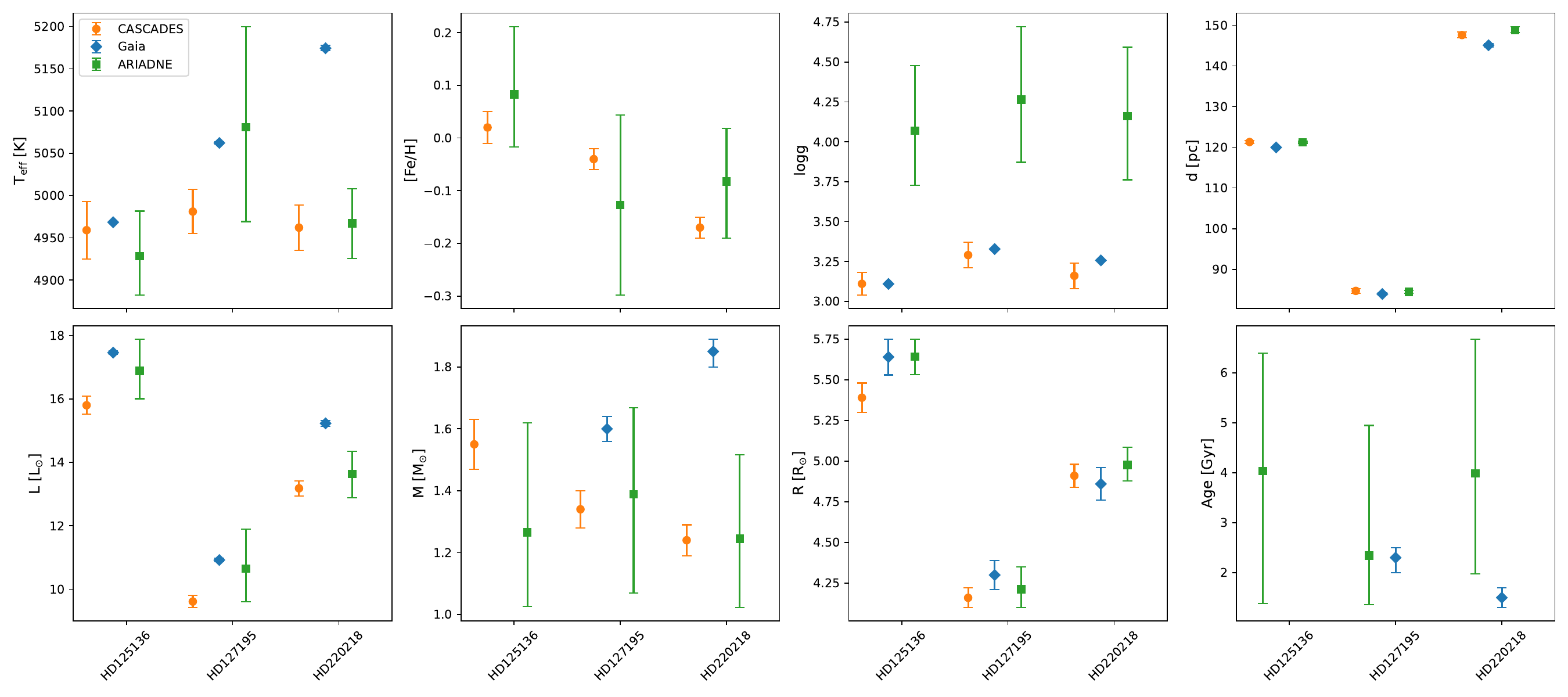}
\caption{Graphical representation of stellar parameters for the three stars and the three determinations methods discussed.}\label{Fig:SteParams}
\centering
\end{figure*}

\section{CORALIE files header keywords used for quality control}\label{App:CORALIEkws}

The CORALIE \texttt{fits} files contain a large number of automatically generated keywords, of which several are particularly useful for quality control:

\begin{itemize}
    \item \texttt{HIERARCH ESO DRS DRIFT QC}: quality control flag for the simultaneous drift calculation; \texttt{PASSED} if the drift is successfully measured and its absolute value is below 110\,m\,s$^{-1}$, otherwise \texttt{FAILED}.
    \item \texttt{HIERARCH ESO OBS MOON SEP} - angular separation, in degrees, between the Moon and the target on the sky (approximate value, computed at the middle of the night).
    \item \texttt{HIERARCH ESO OBS TARG VPICMOON} - difference, in km\,s$^{-1}$, between the RV of the Sun as reflected on the Moon and the (user-provided) approximate RV of the target star. 
\end{itemize}

The \texttt{DRIFT QC} includes two different conditions. First, it indicates whether the drift calculation was successful, with the data satisfying a number of internal quality criteria (e.g. flux thresholds and algorithmic convergence). Second, it ensures that the absolute value of the measured drift is smaller than 110\,m\,s$^{-1}$. 
The drift correction relies on a first-order approximation of variations in the wavelength solution, which is valid only for small perturbations. An empirical analysis of drift amplitudes and correction performance shows that this approximation remains valid below a threshold of 110\,m\,s$^{-1}$. Above this value, the correction becomes unreliable and inflates the RV uncertainties in a manner that is difficult to quantify, in which case a new wavelength solution is required.

The \texttt{MOON SEP} keyword represents the angular separation between the target and the Moon on the sky and is used to assess the risk of moonlight contamination of the spectra. Such contamination is expected to be significant only for separations smaller than $\sim$30$^\circ$, a condition that is not met for our observations. The \texttt{TARG VPICMOON} keyword measures the expected velocity separation between the lunar contamination spectrum and the stellar spectrum and is used to evaluate whether moonlight contamination may significantly affect the measured RVs.

\FloatBarrier

\section{\texttt{kima} analysis}\label{App:kima}

\subsection{Parameter priors}\label{App:priors}

We used weakly informative priors as much as possible, using only information derived from simple summary statistics of the data. The modified log-uniform distribution is a log-uniform distribution with parameters knee, $a$, and upper limit, $b$ \citep{2005ApJ...631.1198G}. Using the modified version allows us to extend the prior to zero, assigning a constant non-zero probability to values below the knee value. 
For the planet semi-amplitude we use a modified log-uniform prior with a knee at 5\,m\,s$^{-1}$ and an upper limit of 200\,m\,s$^{-1}$.

For the period, we used a log-uniform prior from 1 to 5000\,d. The eccentricity is represented with a \cite{1980JHyd...46...79K} distribution, which is an approximation to the $\beta$ distribution proposed in \cite{2013MNRAS.434L..51K}. For the mean anomaly and argument of periastron, we used uniform distributions between 0 and 2$\pi$.

For the stellar systemic RV and the instrument-related parameters, we derive the priors from the datasets: the systemic RV is taken from a uniform distribution between the RV extreme values, and the jitter from a uniform distribution between 5\,m\,s$^{-1}$ to 50\,m\,s$^{-1}$. The offset is drawn from a uniform distribution between minus and plus the RV span. The linear and quadratic slopes are zero-centred normal distributions with a standard deviation set by the data dispersion. All priors are listed in Table\,\ref{Tab:kimapriors}. 

\setlength{\extrarowheight}{3pt}

\begin{table}
\centering

\caption{Priors used in the \texttt{kima} analysis.} 
\label{Tab:kimapriors}

\begin{tabular}{ l c c c }

\hline \hline

\multirow{2}{*}{Group} & \multirow{2}{*}{Parameter} & \multirow{2}{*}{Prior} & \multirow{2}{*}{\small{units}} \\
 & & &  \\

\hline 

\multirow{4}{*}{Planet} & $K_p$ & $\mathcal{MLU}$(5,\,200) & m\,s$^{-1}$ \\[2pt]
 & $P$ & $\mathcal{LU}$(1.0, $\Delta t$) & d \\[2pt]
 & $e$ & $\mathcal{K}(0.867,\,3.03)$ & -- \\[2pt]
 & $\phi$, $\omega$ & $\mathcal{U}(0,\,2\pi)$ & -- \\[2pt]

 \hline

\multirow{2}{*}{Instruments}  & Jitter & $\mathcal{U}(0.05,\,5)$ & \multirow{2}{*}{m\,s$^{-1}$} \\[2pt]
& Offset & $\mathcal{U}(-\Delta RV, \Delta RV)$ & \\[2pt]

\hline

\multirow{3}{*}{Star} & Linear slope & $\mathcal{N}$(0,\,10$^{\frac{\Delta RV}{\Delta t}}$) & \multirow{3}{*}{m\,s$^{-1}$} \\[2pt]
 & Quadratic slope & $\mathcal{N}$(0,\,10$^{\frac{\Delta RV}{\Delta t^2}}$) & \\[2pt]
 & Systemic RV & $\mathcal{U}(RV_\mathrm{min}, RV_\mathrm{max})$ & \\[2pt]
 
\hline
\end{tabular}
\end{table}

\setlength{\extrarowheight}{2pt}

\subsection{Corner plots}\label{App:corners}

\begin{figure*}
\center
\includegraphics[width=\textwidth]{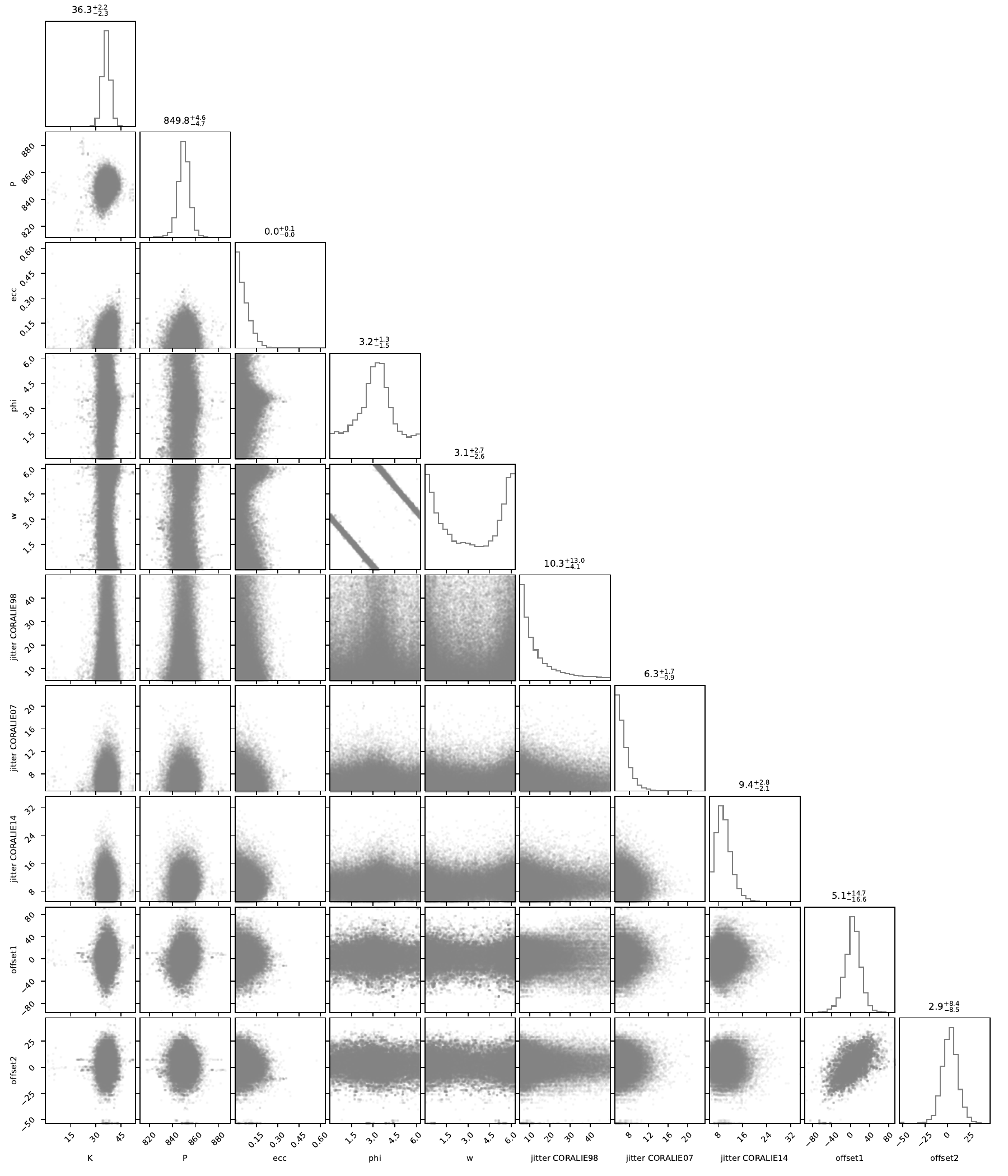}
\caption{Corner plot of the parameters posterior distributions for the RV modelling with \texttt{kima} for HD\,125136.}\label{Fig:cornerHD125136}
\centering
\end{figure*}

\begin{figure*}
\center
\includegraphics[width=\textwidth]{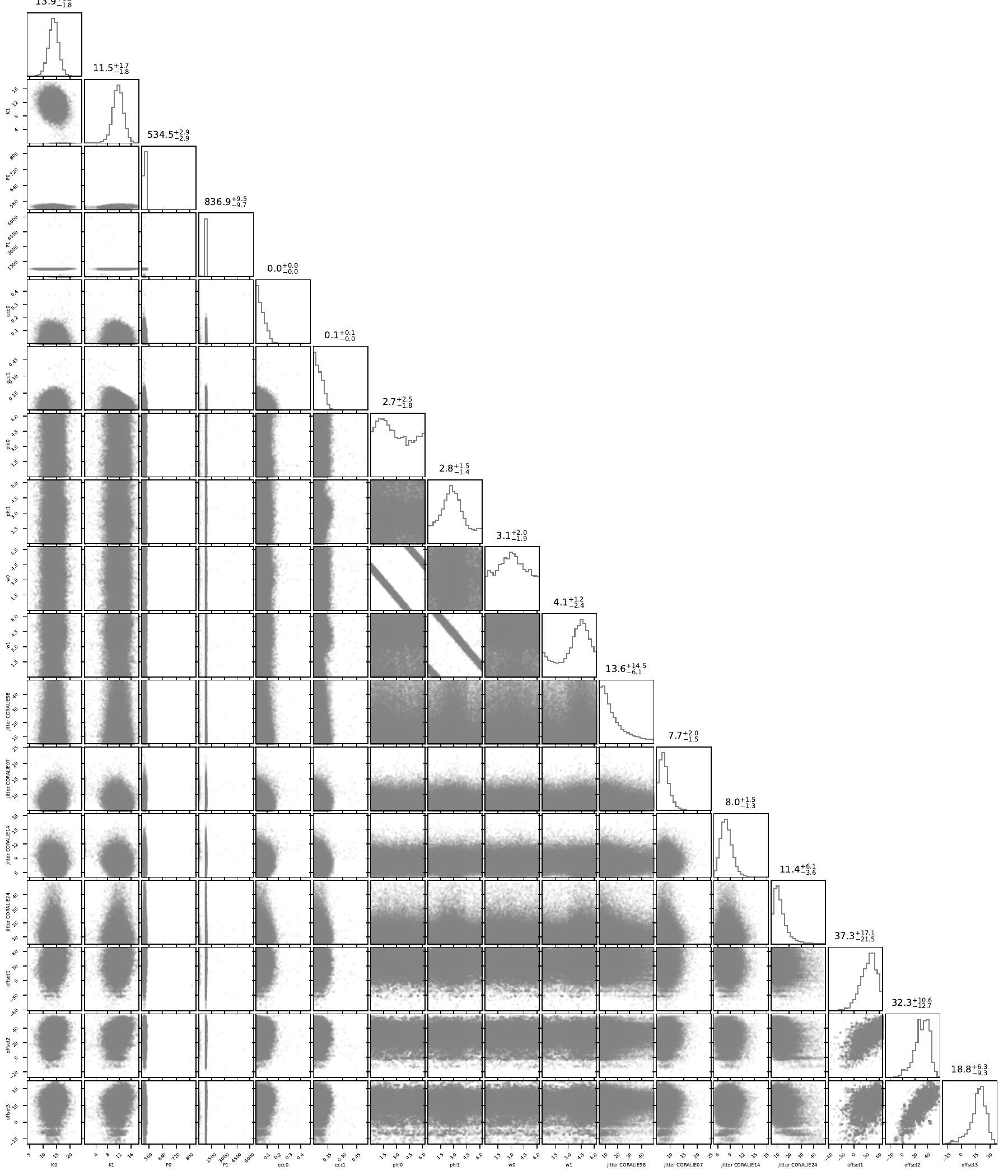}
\caption{Same as Fig\,\ref{Fig:cornerHD125136} but for HD\,127195.}\label{Fig:cornerHD127195}
\centering
\end{figure*}

\begin{figure*}
\center
\includegraphics[width=\textwidth]{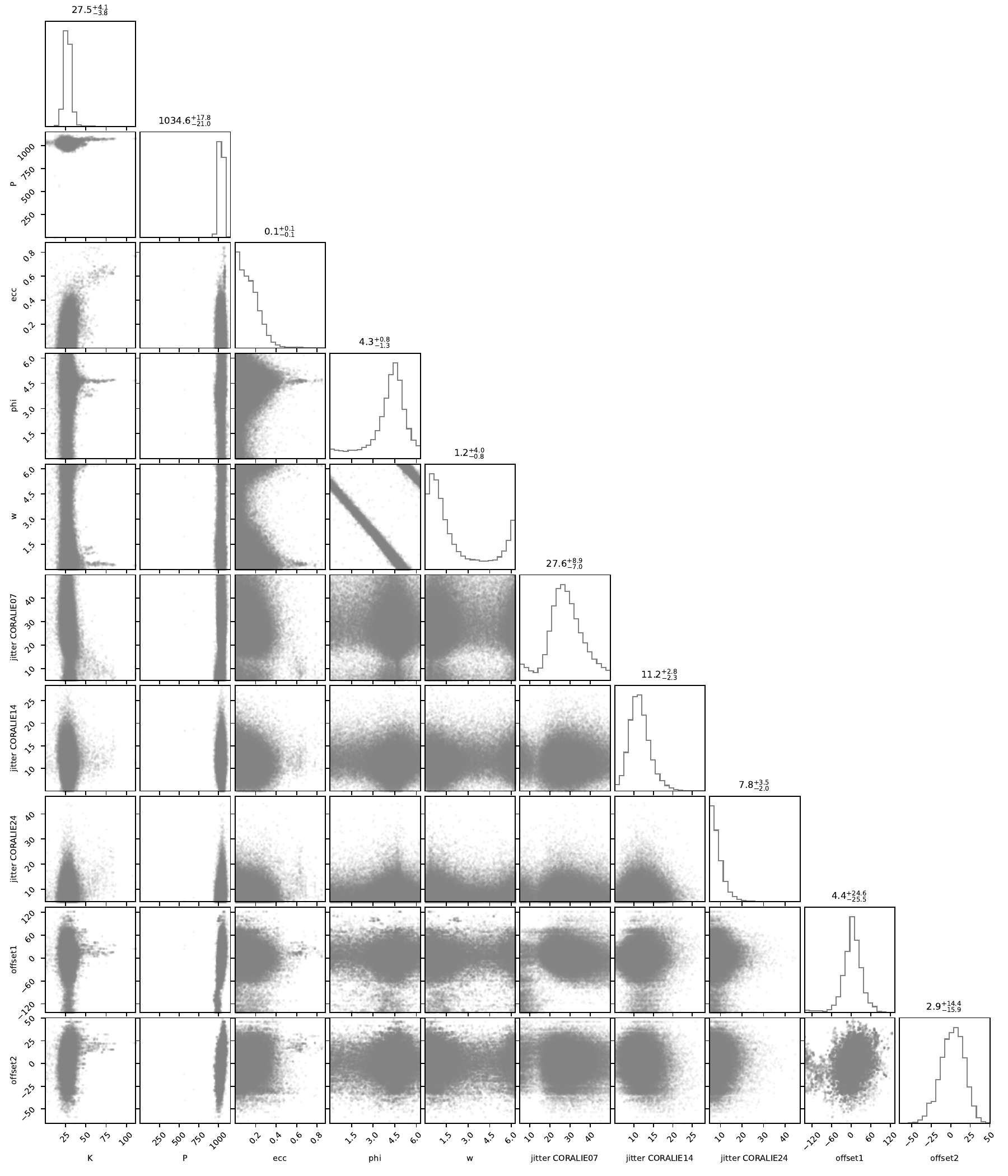}
\caption{Same as Fig\,\ref{Fig:cornerHD125136} but for HD\,220218.}\label{Fig:cornerHD220218}
\centering
\end{figure*}

\FloatBarrier

\section{Pulsation campaign}\label{App:pulsres}

\begin{table}
\centering
\caption{Parameters for the oscillation-averaging campaign on HD\,127195.}
\label{Tab:puls}
\begin{tabular}{lccc}
\hline\hline
Parameter$^\dagger$ & Unit & HD\,127195\,b & HD\,127195\,c \\[3pt]
\hline

$K$      & \tiny{[m\,s$^{-1}$]} & $13.73_{-1.98}^{+1.89}$ & $11.61_{-1.87}^{+1.75}$ \\
$P$      & \tiny{[d]}           & $533.71_{-3.04}^{+3.14}$ & $835.60_{-9.96}^{+10.07}$ \\
$e$      & \tiny{--}            & $0.04_{-0.03}^{+0.05}$   & $0.05_{-0.04}^{+0.05}$ \\
$\phi$   & \multirow{2}{*}{\tiny[$^\circ$]} & $2.69_{-1.86}^{+2.60}$ & $2.92_{-1.56}^{+1.62}$ \\
$\omega$ & & $3.31_{-2.02}^{+1.82}$ & $4.04_{-2.58}^{+1.34}$ \\[3pt]
\hline
jitter$_\texttt{COR98}$     & \multirow{8}{*}{\tiny{[m\,s$^{-1}$]}} & \multicolumn{2}{c}{$14.19_{-6.54}^{+15.24}$} \\
jitter$_\texttt{COR07}$     & & \multicolumn{2}{c}{$7.88_{-1.51}^{+2.06}$} \\
jitter$_\texttt{COR14}$     & & \multicolumn{2}{c}{$8.17_{-1.25}^{+1.53}$} \\
jitter$_\texttt{COR24}$     & & \multicolumn{2}{c}{$11.40_{-3.48}^{+5.77}$} \\
jitter$_\texttt{COR25}$     & & \multicolumn{2}{c}{$7.34_{-1.70}^{+3.77}$} \\
offset$_\texttt{CORE07-98}$ & & \multicolumn{2}{c}{$31.94_{-20.07}^{+17.52}$} \\
offset$_\texttt{COR14-07}$  & & \multicolumn{2}{c}{$29.57_{-13.52}^{+8.61}$} \\
offset$_\texttt{COR24-14}$  & & \multicolumn{2}{c}{$16.32_{-8.26}^{+7.41}$} \\
offset$_\texttt{COR25-24}$  & & \multicolumn{2}{c}{$-3.11_{-7.59}^{+7.49}$} \\[3pt]
\hline
slope & \multirow{2}{*}{\tiny[m\,s$^{-1}$]}  & \multicolumn{2}{c}{$0.002_{-0.002}^{+0.002}$} \\
vsys  & & \multicolumn{2}{c}{$-18.39_{-7.80}^{+10.56}$} \\[4pt]

\hline
\end{tabular}
\tablefoot{
\tablefoottext{$\dagger$}{Expected value provided by the median and uncertainties by the 16--84\% credible intervals of the posterior distribution.}
}
\end{table}

\begin{table}
\centering
\caption{Residual rms and inferred jitter per instrument for the 2025 campaign (values from the MAP solution).}
\label{Tab:rms_jitter_2025}
\begin{tabular}{lcc}
\hline
\hline
\multirow{2}{*}{Instrument} & rms & jitter \\
 & \multicolumn{2}{c}{\small{[m\,s$^{-1}$]}} \\[3pt]
\hline
\small{\texttt{COR98}} & 6.97 & 8.51 \\
\small{\texttt{COR07}} & 8.40 & 7.15 \\
\small{\texttt{COR14}} & 8.76 & 6.78 \\
\small{\texttt{COR24}} & 10.31 & 10.52 \\
\small{\texttt{COR25}} & 5.62 & 5.07 \\
\hline
\end{tabular}
\end{table}

\begin{figure}
\center
\includegraphics[width=0.98\columnwidth]{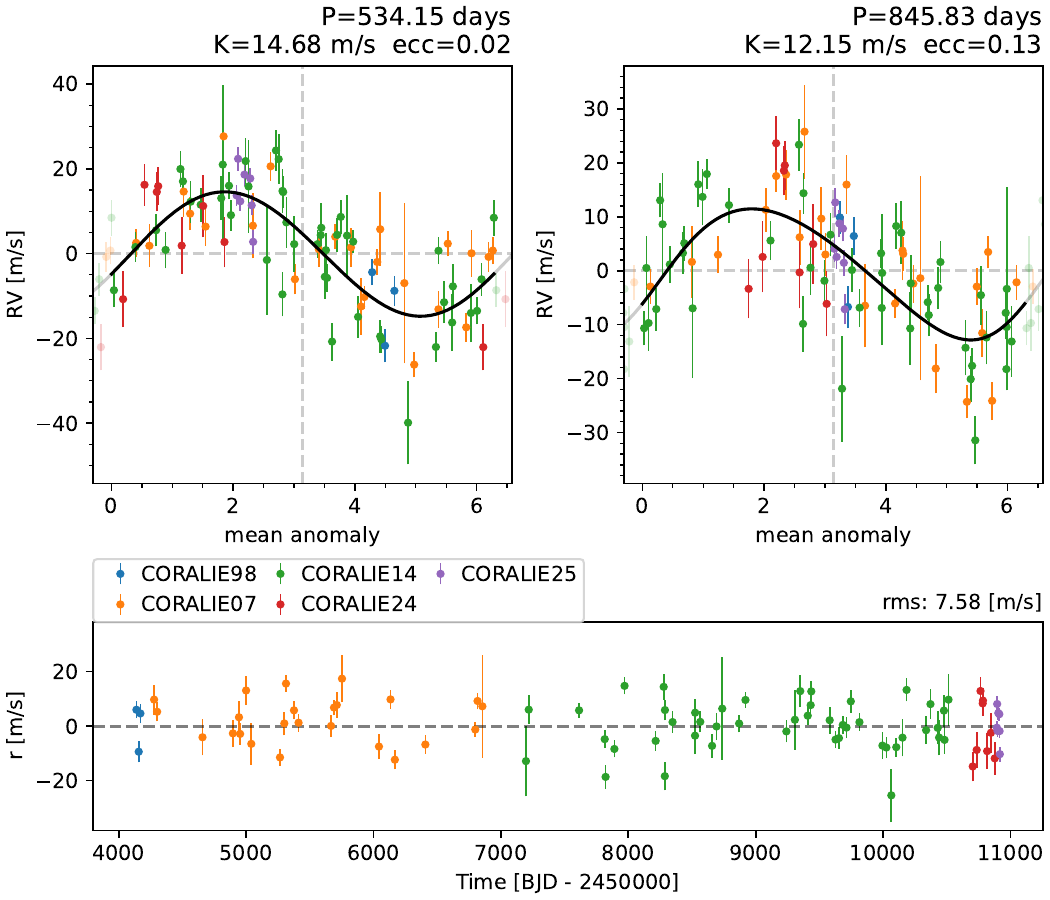}
\caption{Phase plot and residuals as a function of mean anomaly for the two planets detected around HD\,127195 when using pulsation-reducing campaign.}\label{Fig:phaseHD127195_Camp25}
\centering
\end{figure}

\end{appendix} 

\end{document}